\documentclass{aa}
\pdfoutput=1

\usepackage{savesym}
\usepackage{url}
\usepackage{multirow}
\usepackage{bigdelim}
\usepackage{hyperref}
\usepackage[version=3]{mhchem}
\usepackage{txfonts}
\restoresymbol{TXF}{iint}
\usepackage[utf8]{inputenc}
\usepackage{natbib}
\bibpunct{(}{)}{;}{a}{}{,}
\usepackage{pgfplots}
\pgfplotsset{compat=newest}
\pgfplotsset{ tick scale binop=\times }
\usepgfplotslibrary{external}
\usepackage{fixltx2e}
\usepackage{pdflscape}

\newcommand{\ratran}{{\it ratran}}
\newcommand{\herschel}{{\it Herschel}}
\newcommand{\machholz}{C/2004 Q2 (Machholz)}
\newcommand{\parents}{\ce{CH3OH}, HCN, \ce{H^{13}CN}, HNC, and CO}
\newcommand{\detected}{\ce{CH3OH}, HCN, \ce{H^{13}CN}, HNC, \ce{H2CO}, CO, and CS}
\newcommand{\sure}{\ce{CH3OH}, HCN, HNC, \ce{H2CO}, CO, CS}
\newcommand{\molecules}{\ce{H2O}, \parents{}}
\newcommand\kms{\ifmmode{\rm km\thinspace s^{-1}}\else km\thinspace s$^{-1}$\fi}
\newcommand\ms{\ifmmode{\rm m\thinspace s^{-1}}\else m\thinspace s$^{-1}$\fi}
\newcommand\s{\ifmmode{\rm molec.~s^{-1}}\else molec.~s$^{-1}$\fi}
\newcommand\rottemp{$40 \pm 3$~K}
\newcommand\rott{$85 \pm 7$~K}
\newcommand\cratio{$97 \pm 30$}
\newcommand\vexp{0.75~\kms}
\newcommand{\rh}{r_\mathrm{h}}
\newcommand{\xne}{x_{n_\mathrm{e}}}

\DeclareGraphicsExtensions{.pdf,.png,.jpg}

\begin{document}

\title{Submillimetric spectroscopic observations of volatiles in comet
\machholz{}\thanks{Based on observations carried out with the 10-m
Submillimeter Telescope at the Arizona Radio Observatory,
Steward Observatory, Mount Graham, Arizona, USA.}}

\author{M.~de~Val-Borro\inst{\ref{inst1}}\fnmsep\thanks{{\it Current address}:
      Department of Astrophysical Sciences, Princeton University, NJ 08544, USA}
	\and P.~Hartogh\inst{\ref{inst1}}
	\and C.~Jarchow\inst{\ref{inst1}}
	\and M.~Rengel\inst{\ref{inst1}}
	\and G.~L.~Villanueva\inst{\ref{inst2}}\fnmsep\inst{\ref{inst3}}
	\and M.~Küppers\inst{\ref{inst4}}
	\and N.~Biver\inst{\ref{inst5}}
	\and D.~Bockel\'ee-Morvan\inst{\ref{inst5}}
	\and J.~Crovisier\inst{\ref{inst5}}
	}

\titlerunning{Submillimetric observations of comet \machholz{}}
\authorrunning{M.~de~Val-Borro et al.}

\institute{Max Planck Institute for Solar System Research,
  Max-Planck-Str.~2, 37191 Katlenburg-Lindau, Germany\\
    \email{[deval;hartogh;jarchow;rengel]@mps.mpg.de}\label{inst1}
  \and Solar System Exploration Division, NASA Goddard Space Flight
  Center, Greenbelt, MD 20771, USA\\
  \email{geronimo.villanueva@nasa.gov}\label{inst2}
  \and Department of Physics, Catholic University of America,
  Washington, DC 20064, USA\label{inst3}
  \and Rosetta Science Operations Centre, European Space Astronomy
  Centre, European Space Agency, PO Box 78, 28691 Villanueva de la
  Ca\~nada, Madrid, Spain\\
  \email{michael.kueppers@sciops.esa.int}\label{inst4}
  \and LESIA, Observatoire de Paris, CNRS, UPMC, Universit\'e
  Paris-Diderot, 5 place Jules Janssen, 92195 Meudon, France\\
  \email{[nicolas.biver;dominique.bockelee;jacques.crovisier]@obspm.fr}\label{inst5}
}

\date{Received 3 March 2012 / Accepted 4 July 2012}

\abstract
  {Submillimeter spectroscopic observations of comets provide an important tool
  for understanding their chemical composition and enable a taxonomic classification.}
  {We aim to determine the production rates of several parent- and
  product volatiles and the \ce{^{12}C}/\ce{^{13}C} isotopic carbon
  ratio in the long-period comet \machholz{}, which is likely to
  originate from the Oort Cloud.}
  {The  line emission from several molecules in the coma was measured
  with high signal-to-noise ratio in January 2005 at heliocentric
  distance of 1.2 AU by means of high-resolution spectroscopic
  observations using the Submillimeter Telescope (SMT) at the Arizona
  Radio Observatory (ARO).}
  {We have obtained production rates of several volatiles (\detected{})
  by comparing the observed and simulated line-integrated intensities.
  We calculated the synthetic profiles using a radiative transfer code
  that includes collisions between neutrals and electrons, and the
  effects of radiative pumping of the fundamental vibrational levels by
  solar infrared radiation.  Furthermore, multiline observations of the
  \ce{CH3OH} $J$ =  7--6 series allow us to estimate the rotational
  temperature using the rotation diagram technique.  We find that the
  \ce{CH3OH} population distribution of the levels sampled by these
  lines can be described by a rotational temperature of \rottemp{}.
  Derived mixing ratios relative to hydrogen cyanide are
  CO/\ce{CH3OH}/\ce{H2CO}/CS/HNC/\ce{H^{13}CN}/HCN =
  30.9/24.6/4.8/0.57/0.031/0.013/1 assuming a pointing offset of
  8\arcsec\ due to the uncertain ephemeris at the time of the
  observations and the telescope pointing error.}
  {The measured relative molecular abundances in \machholz\ are between 
  low- to typical values of those obtained in Oort Cloud comets,
  suggesting that it has visited the inner solar system previously and
  undergone thermal processing.  The HNC/HCN abundance ratio of $\sim
  3.1$\% is comparable to that found in other comets, accounting for the
  dependence on the heliocentric distance, and could possibly be
  explained by ion-molecule chemical processes in the low-temperature
  atmosphere.  From a tentative \ce{H^{13}CN} detection, the measured
  value of \cratio\ for the \ce{H^{12}CN}/\ce{H^{13}CN} isotopologue
  pair is consistent with a telluric value.  The outgassing variability
  observed in the HCN production rates over a period of two hours is
  consistent with the rotation of the nucleus derived using different
  observational techniques.
  }


\maketitle

\section{Introduction}\label{sec:intro}

Comets spend most of their lifetime in the outer solar system and therefore
have not undergone much thermal processing.  Line emission from
cometary atmospheres at submillimeter and radio wavelengths is a very useful
tool for studying their physical and chemical conditions and relation with other
bodies in the solar system \citep{2002EM&P...90..323B,2004come.book..391B}.
The coma structure and expansion velocity can be derived by fitting the
observed line shapes using a molecular excitation code.  In addition, mixing
ratios of volatiles such as \ce{CH3OH}, CO and CS can be compared with observed
chemical abundances in protoplanetary disks to improve our understanding of
planet formation processes.

The composition of comets has been investigated in the last two decades
to develop a classification based on abundances of primary chemical
species that displays a great compositional diversity
\citep{1995Icar..118..223A,2003AdSpR..31.2563M,2004come.book..391B}.
More than 20 parent volatile species that release directly from ices in
the nucleus, in addition to several radicals and ions formed by
photodissociation in the coma, have been detected via ground-based
spectroscopic surveys at infrared and submillimeter wavelengths and
in situ measurements.  The composition of some cometary ices show strong
evidence of processing in the solar nebula and can provide clues about
their place of formation and subsequent evolution.  The abundance of HCN
relative to water has been observed to be roughly constant with a value
of 0.1\% in several comets for a wide range of heliocentric distances
\citep{2002EM&P...90..323B}.  Other species show a wide spread of
production rates. For instance, \ce{CH3OH} has been found to have a
variable abundance relative to water, ranging between less than 0.15\% to
6\% at different heliocentric distances \citep{2004come.book..391B}.
There is no evident correlation between the observed relative abundances
and the dynamical class of the comets.  Water is the most dominant
volatile species and is typically used to determine relative abundances.
Although water is not directly accessible from the ground at
submillimeter wavelengths, it has been observed from space using the
Submillimeter Wave Astronomical Satellite (SWAS), Odin and \herschel\
satellites
\citep{2000ApJ...539L.151N,2003A&A...402L..55L,2009P&SS...57.1596H} or
inferred by observations of the hydroxyl (OH) radical at radio
wavelengths \citep{2002A&A...393.1053C}.  \herschel{}'s Heterodyne
Instrument for the Far Infrared (HIFI) is able to determine water
production rates accurately \citep{2010A&A...518L.150H}. Direct
measurements of water can also be performed using ground-based
telescopes by observations of non-resonance fluorescence emission at
infrared wavelengths \citep{1986Sci...232.1523M,2009ApJ...699.1563B}.

Comet \object{\machholz{}} was discovered on 27 August 2004 by
Donald~E.~Machholz \citep{2004IAUC.8394....1M}.  The comet passed
perihelion on 24 January 2005 at a heliocentric distance of $\rh =
1.205$~AU and geocentric distance of $\Delta = 0.435$~AU.  During its
closest approach to Earth at a distance of 0.35 AU on 6 January 2005 it
reached a naked-eye visual magnitude of $m_\mathrm{v}\sim3.5$ 
as reported in the International Comet Quarterly.  Owing to its favorable
viewing geometry in the northern hemisphere, the comet was extensively
observed from the ground at various wavelengths.  \machholz{} is a
long-period highly eccentric comet whose origin is most likely the Oort
Cloud according to the comet classification scheme by
\citet{1996ASPC..107..173L}, with an approximate initial orbital period of
118\,000~years and eccentricity of 0.9994658 before the comet was
perturbed gravitationally in the inner solar system \citep{2004U31,
NK1352}.  These values are close to the perihelion osculating elements
\citep[see][JPL Small-Body
Database\footnote{\url{http://ssd.jpl.nasa.gov/sbdb.cgi?sstr=C/2004+Q2}}]{2005MPC..54558...1M}.
Its dynamical classification is still a matter of debate because of
the strong non-gravitational forces that make a backward orbital
integration very unreliable.

In this paper we present high-resolution spectroscopic observations of
several volatiles from comet \machholz{} acquired at the Submillimeter
Telescope (SMT).  Seven species are detected, namely \sure{}, and a
marginal detection of \ce{H^{13}CN}. The comet was observed shortly
pre-perihelion in January 2005 when it was at a distance $\sim 0.36$~AU
from Earth.  These observations provide information about the outgassing
of several molecules and an isotopologue of HCN relative to HCN, which
is often used as a proxy for water in cometary taxonomies.  We calculate
the \ce{CH3OH} rotational temperature of the ground vibrational level
from several rotational lines and production rates for the observed
molecules using a radiative transfer code to fit the observed line
intensities.  Section~\ref{sec:observations} presents our SMT
observations of comet \machholz{} and the reduction method.  In
Sect.~\ref{sec:results} the radiative transfer models and analysis of
the observations are described.  Finally, we discuss the obtained
results in Sect.~\ref{sec:discussion}.

\section{Observations}\label{sec:observations}

A spectral line survey of primary volatile species in comet \machholz{}
was made using the SMT telescope located at the Mount Graham
International Observatory (MGIO), a division of Steward Observatory on
Mount Graham, Arizona \citep{1996RvMA....9..111B,1999PASP..111..627B}.
The SMT has a parabolic 10-m primary dish and a hyperbolic secondary
reflector.  The observations were performed with the 0.8 mm
double-sideband receiver using various spectrometers
\citep{2005DPS....37.1108V,2011AOGS}.  Detected emission lines may have
a different gain response depending upon which sideband the lines were
observed in.  Thus, the sideband gain ratio
($G_\mathrm{USB}/G_\mathrm{LSB}$) deviates from unity and introduces an
additional uncertainty of $\sim$ 10\% in the absolute brightness
temperature calibration.  A typical system temperature of 150 K was
attained during the observations.  We observed the comet simultaneously
with the chirp transform spectrometer (CTS) with a bandwidth of 215~MHz,
and the acousto-optical spectrometers (AOSA, AOSB and AOSC) with total
bandwidths of 1~GHz, 970~MHz and 250~MHz, respectively.  The spectral
resolution provided was 40 kHz for the CTS, and 934, 913 and 250 kHz for
the AOSA, AOSB and AOSC.  One of the purposes of the observations was to
test the performance of the newly installed high-resolution CTS built at
the Max Planck Institute for Solar System Research
\citep{1990MeScT...1..592H,2004ExA....18...77V,2006ITMTT..54.1415V}.
High spectral resolution is crucial for resolving the shape of rotational
lines in comets and study the gas velocity and asymmetries related to
non-isotropic outgassing.

\begin{table*}
  \caption{Log of the SMT observations of comet \machholz{} in January
  2005. 
  }
  \label{tbl:log}
  \centering
  \begin{tabular}{c c r@{}l c c c c c c}
    \hline\hline
    Date\tablefootmark{a} & Molecule & \multicolumn{2}{l}{Transition}&
    $N$\tablefootmark{b} & Integration\tablefootmark{c} &
    $\langle\rh\rangle$\tablefootmark{d}& $\langle\Delta\rangle$\tablefootmark{e}&
    $\langle\phi\rangle$\tablefootmark{f} & Beam size\\
    (UT) & & & & & (min) & (AU) & (AU) & & \\
    \hline
    \multirow{2}{*}{12.96} \ldelim\{{2}{0mm}[] & HCN           & 4--      & 3                            & \multirow{2}{*}{13} & \multirow{2}{*}{107.2} & \multirow{2}{*}{1.220} & \multirow{2}{*}{0.361} & \multirow{2}{*}{42\fdg5} & 20\farcs93 \\
                                               & \ce{H2CO}     & 4--      & 3                            &                     &                        &                        &                        &                          & 21\farcs09 \\
    \multirow{4}{*}{13.20} \ldelim\{{4}{0mm}[] & CO            & 3--      & 2                            & \multirow{4}{*}{33} & \multirow{4}{*}{270.6} & \multirow{4}{*}{1.219} & \multirow{4}{*}{0.362} & \multirow{4}{*}{42\fdg6} & 21\farcs46 \\
                                               & CS            & 7--      & 6                            &                     &                        &                        &                        &                          & 21\farcs64 \\
                                               & \ce{H^{13}CN} & 4--      & 3                            &                     &                        &                        &                        &                          & 21\farcs49 \\
                                               & \ce{CH3OH}    & $13_1$-- & $13_{0}$A\textsuperscript{-} &                     &                        &                        &                        &                          & 21\farcs65 \\
    13.95                                      & \ce{CH3OH}    & 7--      & 6                            & 16                  & 133.0                  & 1.217                  & 0.365                  & 43\fdg3                  & 21\farcs93 \\
    14.17                                      & HNC           & 4--      & 3                            & 22                  & 183.0                  & 1.217                  & 0.366                  & 43\fdg5                  & 20\farcs46 \\
    15.14                                      & HNC           & 4--      & 3                            & 39                  & 329.2                  & 1.215                  & 0.371                  & 44\fdg2                  & 20\farcs46 \\
    \multirow{4}{*}{16.16} \ldelim\{{4}{0mm}[] & CO            & 3--      & 2                            & \multirow{4}{*}{47} & \multirow{4}{*}{378.0} & \multirow{4}{*}{1.213} & \multirow{4}{*}{0.376} & \multirow{4}{*}{45\fdg0} & 21\farcs46 \\
                                               & CS            & 7--      & 6                            &                     &                        &                        &                        &                          & 21\farcs64 \\
                                               & \ce{H^{13}CN} & 4--      & 3                            &                     &                        &                        &                        &                          & 21\farcs49 \\
                                               & \ce{CH3OH}    & $13_1$-- & $13_{0}$A\textsuperscript{-} &                     &                        &                        &                        &                          & 21\farcs65 \\
    \hline
  \end{tabular}
  \tablefoot{In some receiver tunings several lines were observed
  simultaneously in the lower and upper sidebands.
  \tablefoottext{a}{Start times in universal time (UT) fractional day values.}
  \tablefoottext{b}{Number of individual scans.}
  \tablefoottext{c}{Total integration time.}
  \tablefoottext{d}{Heliocentric distance.}
  \tablefoottext{e}{Geocentric distance.}
  \tablefoottext{e}{Solar phase angle (Sun--\machholz{}--Earth).}
  }
\end{table*}

Comet \machholz{} was observed near perihelion during six nights in
the period 13--18 UT January 2005.  Here we focus on the observations
acquired on the first four nights with good observing conditions. The
data were taken using the standard position switching observing mode,
where a reference sky position separated from the comet by 0\fdg5 was
observed for the same amount of time and subtracted from the on-source
observations.  We used a 30-second exposure for the source and
reference positions with 8-minute scans.  A summary of the observing log
with the total integration times for each line is shown in
Table~\ref{tbl:log}.

The main beam equivalent brightness temperature scale, $T_\mathrm{mB}$,
was corrected for the beam efficiency of the telescope estimated from
observations of Mars and Saturn in the interval fall 2007--spring 2008%
\footnote{\url{http://kp12m.as.arizona.edu/smt_docs/smt_beam_eff.htm}}
and calibrated using the chopper-wheel method
\citep{1976ApJS...30..247U}. The frequency scale was converted into
Doppler velocities in the nucleus frame and corrected for the relative
motion of the comet with respect to the observer.  We used orbital
elements provided by the JPL HORIZONS
system\footnote{\url{http://ssd.jpl.nasa.gov/?horizons}} to track the
comet and the ephemeris to calculate the position and relative motion of
the comet with respect to the telescope.  Our observing strategy
included pointing and calibration observations approximately every hour.
The pointing of the telescope was checked and corrected by dedicated
reference observations of bright sources like Saturn, the protobinary
system W3(OH) and the massive star-forming region DR21(OH) because they were
close to the comet.  Typical pointing errors were $\sim 4\arcsec$, i.e.,
about a quarter of the half-power beam width (HPBW) at the observed
frequencies.

These observations were very challenging owing to the relatively
uncertain ephemeris of \machholz{} since they were carried out about
five months after the comet was discovered, and particularly because of
the strong non-gravitational forces needed to reproduce the observations
afterwards.  Its orbit changed significantly during its passage through
the inner solar system due to planetary perturbations from an initial
orbital parameter of $a_0^{-1} = 0.000404$ AU\textsuperscript{-1} to a
future value of 0.001856 AU\textsuperscript{-1} \citep{2004U31}.  At the
time of the observations, the comet was moving with a fast apparent
motion of $\sim 300$\arcsec\ per hour in the sky relative to the
background stars. This introduced an additional term of uncertainty in
the pointing accuracy.  Comparing the ephemeris actually used for the
tracking with the HORIZONS ephemeris calculated from the after-the-fact
orbit including non-gravitational accelerations, we obtain an average
pointing offset of 4\arcsec.  Including the effect of the telescope
pointing error, the offset was at maximum 8\arcsec.  This value was
considered in our computation of the production rates for the observed
species shown in Table~\ref{tbl:mixing}.

\begin{table*}
  \caption{Comparison of production rates of detected species relative
  to HCN and \ce{H2O} in comet \machholz{} with statistical
  uncertainties.}
  \label{tbl:mixing}
  \centering
  \begin{tabular}{c c c c c c c c}
    \hline\hline
    Molecule & & $Q$\tablefootmark{a} & $Q/Q_\ce{HCN}$ &
    $Q/Q_\ce{H2O}$\tablefootmark{b} & $Q/Q_\ce{H2O}$\tablefootmark{c}  &
    $Q/Q_\ce{H2O}$\tablefootmark{d}  \\
    & & (\s) & & (\%) & (\%) & (\%) \\
    \hline
    HCN                         &                     & $(2.26 \pm 0.02) \times 10^{26}$                                             & $1$                             & $0.084 \pm 0.001$                  & $0.15^{+0.01}_{-0.02}$                                      &  $0.16 \pm 0.01$                                             \\
    CO                          &                     & $(7.0 \pm 0.6) \times 10^{27}$                                               & $30.9 \pm 2.6$                  & $2.6 \pm 0.2$                      & $5.07 \pm 0.51$                                             \\
    \multirow{2}{*}{\ce{CH3OH}} &                     & \multirow{2}{*}{$(5.5 \pm 0.6) \times 10^{27}$}                              & \multirow{2}{*}{$24.6 \pm 2.5$} & \ldelim\{{2}{12mm}[$2.1 \pm 0.2$ ] & $2.14 \pm 0.12$                                             &  $1.2 \pm 0.1$                                               \\
                                &                     &                                                                              &                                 &                                    & \phantom{\tablefootmark{e}}$1.54 \pm 0.07$\tablefootmark{e} &  \phantom{\tablefootmark{e}}$1.65 \pm 0.09$\tablefootmark{e} \\
    \multirow{2}{*}{\ce{H2CO}}  & \ldelim\{{2}{0mm}[] & $(3.87 \pm 0.20) \times 10^{26}$                                             & $1.7 \pm 0.1$                   & $0.14 \pm 0.01$                    & \multirow{2}{*}{$0.11 \pm 0.03$}                            &  \multirow{2}{*}{$0.18 \pm 0.01$}                            \\
                                &                     & \phantom{\tablefootmark{f}}$(1.09 \pm 0.06) \times 10^{27}$\tablefootmark{f} & $4.8 \pm 0.3$                   & $0.41 \pm 0.02$                    &                                                             &                                                              \\
    \multirow{2}{*}{CS}         & \ldelim\{{2}{0mm}[] & $(1.15 \pm 0.04) \times 10^{26}$                                             & $(5.1 \pm 1.7)\times10^{-1}$    & $(4.3 \pm 0.1)\times10^{-2}$       &                                                             \\
                                &                     & \phantom{\tablefootmark{f}}$(1.29 \pm 0.04) \times 10^{26}$\tablefootmark{f} & $(5.7 \pm 1.9)\times10^{-1}$    & $(4.8 \pm 0.2)\times10^{-2}$       &                                                             \\
    HNC                         &                     & $(7 \pm 2) \times 10^{24}$                                                   & $(3.1 \pm 0.9)\times10^{-2}$    & $(2.6 \pm 0.7)\times10^{-3}$       &                                                             \\
    \ce{H^{13}CN}               &                     & $(3 \pm 1) \times 10^{24}$                                                   & $(1.3 \pm 0.5)\times10^{-2}$    & $(1.1 \pm 0.4)\times10^{-3}$       &                                                             \\
    \hline
  \end{tabular}
  \tablefoot{
  \tablefoottext{a}{Weighted mean of production rates measured at the
  SMT on 13--16 January 2005 derived from excitation and radiative
  transfer models assuming a pointing offset of 8\arcsec{}.}
  \tablefoottext{b}{A water production rate of $Q_\ce{H2O} = 2.7 \times
  10^{29}\ \s$ was used to derive the mixing ratios, intermediate
  between the measurement on 20 January 2005 by the Odin satellite
  \citep{2007P&SS...55.1058B} and the value derived from observations at
  infrared wavelengths on 19 January 2005
  \citep{2009ApJ...699.1563B}.}
  \tablefoottext{c}{Weighted mean of mixing ratios obtained on 28--29
  November 2004 and 19 January 2005 with NIRSPEC on the Keck II 10-m
  telescope \citep{2009ApJ...699.1563B}}
  \tablefoottext{d}{Mixing ratios obtained on 30 January 2005 with NIRSPEC
  on the Keck II 10-m telescope \citep{2009ApJ...703..121K}}
  \tablefoottext{e}{Revised \ce{CH3OH} mixing ratios from \citet{2012ApJ...747...37V}.}
  \tablefoottext{f}{Production rates for \ce{H2CO} and CS assume the
  presence of an extended source described in the text.}
  }
\end{table*}

The data analysis was performed using the CLASS analysis software, which
is part of GILDAS
package\footnote{\url{http://www.iram.fr/IRAMFR/GILDAS}}, in conjunction
with the NumPy and SciPy libraries of high-level mathematical tools
\citep{scipy,10.1109/MCSE.2007.58}.  A standing wave appears as a
baseline ripple in some of the spectra.  We determined the baseline by
fitting a polynomial to the emission-free background and subtracted it
from the original spectrum.  After removal of the standing wave, the
individual scans were averaged to increase the signal-to-noise ratio
(S/N).  Line intensities were calculated from the weighted averages of
the spectra where the statistical weights are the inverse square of the
root mean square (rms) noise of each individual spectrum.

The comet's heliocentric distance was in the range $\rh =
1.213$--$1.220$ AU, and the geocentric distance was $\Delta =
0.361$--$0.376$ AU during the observing period (see
Table~\ref{tbl:log}).  The HPBW of the telescope varied between
20\farcs5 and 22\arcsec\ at different frequencies, corresponding to
6\,600 and 6\,900 km projected on the comet, so that molecules from the
outer coma contribute to the detected emission.  The line excitation in
this region is dominated by collisions between neutrals and electrons,
and infrared fluorescence by solar radiation.  During the observing run
the solar phase angle (Sun--\machholz{}--Earth) ranged from $42\fdg5$ to
$45\fdg0$.  We detected seven species (\detected{}), whose spectra are
shown in Figs.~\ref{fig:ch3oh}--\ref{fig:ch3oh2} and
\ref{fig:hcn}--\ref{fig:cs}.  The observations were focused on the study
of the hydrogen cyanide chemistry with the detection of HCN, HNC and a
tentative detection of \ce{H^{13}CN} to determine the isotopic
\ce{^{12}C}/\ce{^{13}C} ratio, as well as other volatile species that
sublimate directly from the nucleus into the coma and abundant daughter
species.

In Table~\ref{tbl:mixing} we show a comparison of the mixing ratios of
the detected species with respect to \ce{H2O} with those derived from
infrared observations on similar dates
\citep{2009ApJ...699.1563B,2009ApJ...703..121K}.  For species that are
detected in several nights during our observing run, the mixing ratios
refer to the weighted average of the observed spectra.
Table~\ref{tbl:obs} shows the molecules and transitions detected in our
survey with 1-$\sigma$ uncertainties.  Line frequencies were obtained
from the latest online edition of the JPL Molecular Spectroscopy Catalog
\citep{1998JQSRT..60..883P}.  Most transitions were observed with the
higher resolution CTS, which resolves the line shape, and the AOS
backends simultaneously.  Integrated line intensities shown in
Table~\ref{tbl:obs} were obtained integrating over velocities in the
interval \mbox{[-2, 2]}~\kms unless another line overlaps within that
region.  Doppler shifts in the comet rest frame were obtained as the
first moment of the velocity over the same interval ($\sum_i
{T_\mathrm{mB}}_i v_i / \sum_i {T_\mathrm{mB}}_i$ where $v$ is the
Doppler velocity and the index $i$ refers to the channel number). 

\section{Results}\label{sec:results}

\subsection{Radiative transfer modeling}\label{sec:radtran}

We adopted a molecular excitation and radiative transfer model based on
the Sobolev escape probability method, which includes collisional effects
and infrared fluorescence by solar radiation to derive the production
rates
\citep{1987A&A...181..169B,1997PhDT........51B,1999AJ....118.1850B}.  In
the outer coma, solar infrared pumping of vibrational bands followed by
instantaneous spontaneous decay establishes a fluorescence equilibrium.
To test the effect of the infrared excitation, an independent
molecular excitation model based on the publicly available accelerated
Monte Carlo radiative transfer code \ratran{} was used to calculate the
populations of the rotational levels as a function of the distance from
the nucleus for various molecules and line emission in the cometary coma
\citep{2000A&A...362..697H}\footnote{the source code is available from
\ratran{}'s website at
\url{http://www.sron.rug.nl/~vdtak/ratran/frames.html}}.  This code
includes collisional effects with water molecules and electrons, but
neglects the pumping by solar infrared radiation from the ground-state
vibrational level.  We used the one-dimensional spherically symmetric
version of the code following the description outlined in
\citet{2004ApJ...615..531B}, which has been extensively tested and
used to interpret \herschel{} cometary observations \citep[see
e.g.][]{2010A&A...518L.150H,2010A&A...521L..50D,2011Natur.478..218H}.

The radial gas density profiles for parent molecules \molecules{} were
obtained using the standard Haser spherically symmetric distribution for
parent volatiles \citep{1957BSRSL..43..740H}: \begin{equation}
  n_\mathrm{p}(r)= \frac{Q}{4\pi r^2v_\mathrm{exp}}\,
  \exp\left(-\frac{r\beta_\mathrm{p}}{v_\mathrm{exp}}\right),
\end{equation} where $Q$ is the total production rate in molecules
s$^{-1}$.  The photodissociation rate $\beta_\mathrm{p}$ takes into
account the dissociation and ionization of molecules by the solar UV
radiation, $v_\mathrm{exp}$ is the expansion velocity and $r$ is the
nucleocentric distance.  However, second-generation molecules such as
HNC, CS and \ce{H2CO} are not correctly described by this profile.  In
the last two cases, the density profile of daughter species originating
from the photodissociation of an extended parent source in the coma is
given by \citep{2004come.book..523C}
\begin{equation} n_\mathrm{d}(r)= \frac{Q}{4\pi
  r^2v_\mathrm{exp}}
  \frac{\beta_\mathrm{p}}{\beta_\mathrm{d}-\beta_\mathrm{p}}\, \left(
  \exp\left(-\frac{r\beta_\mathrm{p}}{v_\mathrm{exp}}\right) -
  \exp\left(-\frac{r\beta_\mathrm{d}}{v_\mathrm{exp}}\right) \right),
\end{equation} 
where $\beta_\mathrm{p}$ and $\beta_\mathrm{d}$ are the
parent and daughter photodissociation rates.  We obtained the
photodissociative lifetimes for \ce{CH3OH}, HCN, HNC, \ce{H2CO} and CO
from \citet{1994JGR....99.3777C} assuming the quiet-Sun reference
spectrum, and the CS radical lifetime was taken from
\citet{2011A&A...528A.142B}.  The photodissociation rates were scaled by
the heliocentric distance of \machholz{} at the time of the
observations. Deviations from a spherically symmetric Haser distribution
due to anisotropic outgassing were not considered.

Neutrals in cometary atmospheres are excited by collisions with
other molecules and electrons, which are the dominant effects in the
inner coma, and radiative pumping of the fundamental vibrational levels
by the solar infrared flux.  Infrared pumping of vibrational bands by
solar radiation contributes to the excitation in the outer coma where
the gas and electron densities are low \citep{1987A&A...181..169B}. Most
of the detected emission in comet \machholz{} originates from a region
in which molecules are in an excitation state intermediate between
fluorescence and collision-dominated equilibrium.

Collision rates with water and electrons were obtained for several
molecules from the current version of the Leiden Atomic and Molecular
Database\footnote{\url{http://www.strw.leidenuniv.nl/~moldata/}}
\citep[LAMDA;][]{2005A&A...432..369S}.
When collision rates with water were not available, we scaled the
collision rates with molecular hydrogen by the ratio of their molecular
weights.  However, the calculated line intensities of the rotational
transitions in the ground-based vibrational level are weakly dependent
on the collision cross-sections -- of about a few percent.

We used a water production rate of $Q_\ce{H2O} = 2.7 \times
10^{29}\ \s$, close to the Odin satellite measurement on 20 January 2005
of $(2.64 \pm 0.08)\times10^{29}\ \s$ derived from the observation of
the fundamental transition at 557~GHz \citep{2007P&SS...55.1058B} at
$\rh = 1.208$~AU, and to the water production rates of $(2.73 \pm 0.07)
\times 10^{29}\ \s$ and $(2.76 \pm 0.08) \times 10^{29}\ \s$ derived
from observations at infrared wavelengths on 19 January 2005
\citep{2006ApJ...653..774B,2007ApJ...661L..97B,2009ApJ...699.1563B}.
Excitation parameters in the model are the
neutral gas kinetic temperature, which controls the molecular excitation
in the collisional region, and the electron density.  We assumed a gas
kinetic temperature of 60 K and the rotational temperature derived from
the relative line intensities of the $J$ = 7--6 transitions for 
\ce{CH3OH}.  The electron density and temperature profiles from
\citet{1997PhDT........51B} were assumed.  Since the electron density in
the coma is not a fully known quantity, an electron density scaling
factor of $\xne = 0.5$ with respect to the reference profile deduced
from in situ measurements of comet 1P/Halley was used.  The
expansion velocity is assumed to be constant in the coma, with a value
of $\vexp{}$, obtained from the width of the HCN line. This value is
consistent with the shape of several lines observed with the Institut de
Radioastronomie Millim\'etrique (IRAM) 30-m radio telescope at about the
same time (Biver et al.\ in preparation).

The radiative transfer equation was solved by integrating along various
lines of sight through the coma that covered $2.5 \times$HPBW at a given
offset from the nucleus, to obtain the brightness distribution in the
plane of the sky once the level populations were calculated.  The
beam-averaged emission was then calculated at the distance of the comet.

Production rates were calculated comparing observed line-integrated
intensities with the integrated synthetic lines obtained with the
Sobolev escape probability and \ratran{} codes by fitting the production
rate to the observed line intensity to derive the relative
molecular abundances in the nucleus \citep{2004ApJ...615..531B}.
Opacity effects are negligible for all lines, therefore the difference in
the derived production rates in our models arise only from including the
pumping of excited vibrational bands by infrared solar radiation as an
excitation mechanism \citep{1989A&A...216..278B}.
Table~\ref{tbl:mixing} summarizes the mixing ratios relative to water
and hydrogen cyanide in comet \machholz{}.  We show the derived
production rates with statistical uncertainties for the observed
molecules using parent- and daughter density distributions in
Table~\ref{tbl:obs}. Production rates are affected by pointing errors of
up to 8\arcsec\ since the ephemeris used at the time of the observations
was relatively uncertain. Therefore we considered this pointing offset
to compute the production rates. 

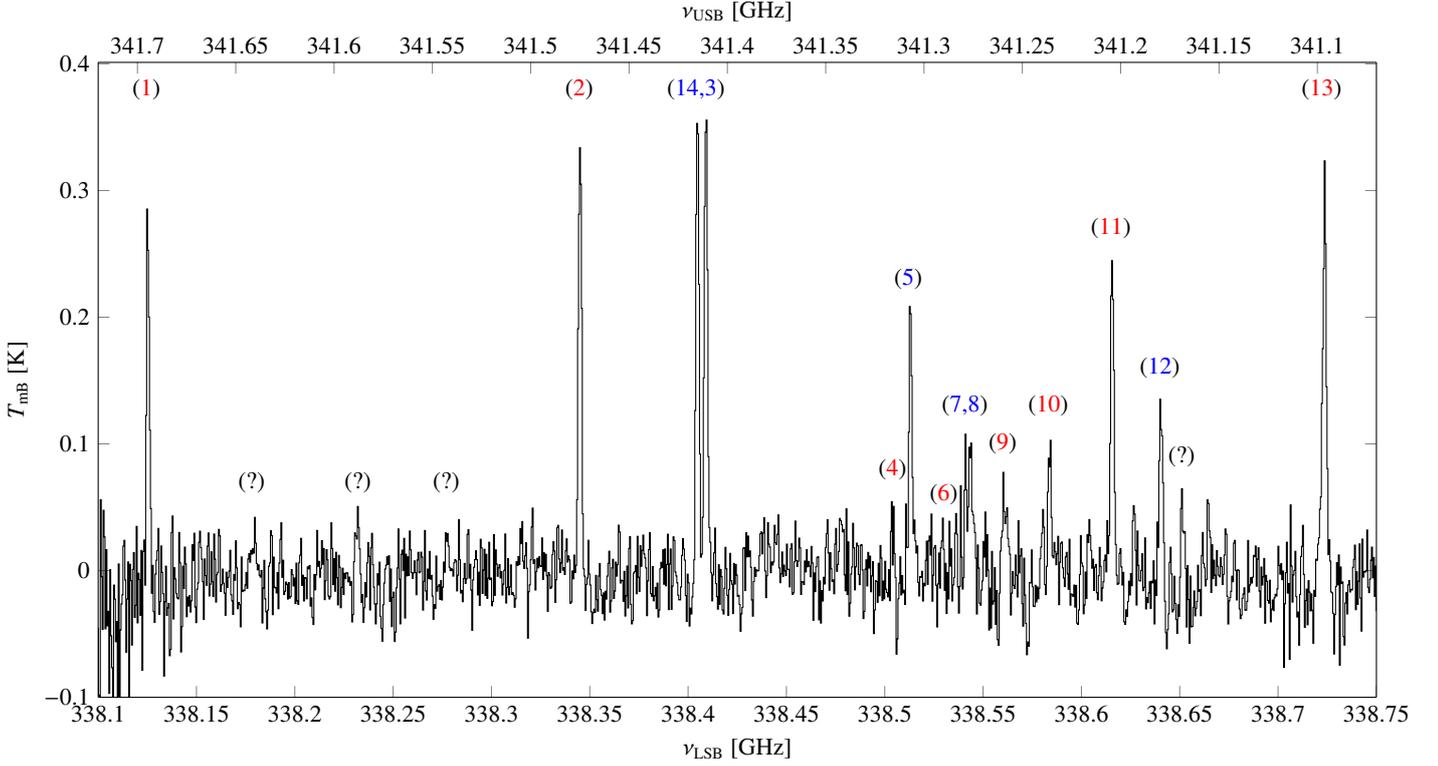
\begin{figure*}
  \centering
  \begin{tikzpicture}
    \begin{axis}[
      xlabel={$\nu_\mathrm{LSB}$ [GHz]},ylabel={$T_\mathrm{mB}$ [K]},
      const plot mark right, enlarge x limits=false,
      xmin=338.1, xmax=338.75, ymin=-0.1,
      axis x line*=bottom,
      width=\hsize, height=10cm]
      \addplot[mark=none] table [x index=0,y index=2]
	{/home/miguel/project/Machholz/data/fits/CH3OH.dat};
      \node at (axis cs: 338.124500, 0.38) {(\textcolor{red}{1})};
      \node at (axis cs: 338.344625, 0.38) {(\textcolor{red}{2})};
      \node at (axis cs: 338.404, 0.38) {(\textcolor{blue}{14,3})};
      \node at (axis cs: 338.504, 0.08) {(\textcolor{red}{4})};
      \node at (axis cs: 338.512, 0.23) {(\textcolor{blue}{5})};
      \node at (axis cs: 338.530, 0.06) {(\textcolor{red}{6})};
      \node at (axis cs: 338.540781, 0.13) {(\textcolor{blue}{7,8})};
      \node at (axis cs: 338.559937, 0.10) {(\textcolor{red}{9})};
      \node at (axis cs: 338.583187, 0.13) {(\textcolor{red}{10})};
      \node at (axis cs: 338.615000, 0.27) {(\textcolor{red}{11})};
      \node at (axis cs: 338.639937, 0.16) {(\textcolor{blue}{12})};
      \node at (axis cs: 338.722218, 0.38) {(\textcolor{red}{13})};
      \node at (axis cs: 338.178, 0.07) {(?)};
      \node at (axis cs: 338.232, 0.07) {(?)};
      \node at (axis cs: 338.277, 0.07) {(?)};
      \node at (axis cs: 338.651, 0.09) {(?)};
    \end{axis}
    \begin{axis}[
      xlabel={$\nu_\mathrm{USB}$ [GHz]},
      xmin=341.07, xmax=341.72,
      x dir=reverse,
      ymin=0, ymax=1,
      axis x line*=top,
      axis y line=none,
      width=\hsize, height=10cm]
    \end{axis}
  \end{tikzpicture}
  \caption{\ce{CH3OH} averaged spectrum obtained on 13.95~UT January
  2005.  The lower and upper x-axis scales represent the frequency of
  the lower and upper sidebands, respectively.  Labels indicate
  detected \ce{CH3OH} spectral lines listed in Table~\ref{tbl:obs}.
  There is a blend of three A\textsuperscript{+} and
  A\textsuperscript{-} emission lines at 338.513~GHz (label 5). The
  observed line at 338.722~GHz is a blend of two E components (label
  13).  The A\textsuperscript{-} line at 341.416~GHz (label 14) is in
  the upper sideband close to the A\textsuperscript{+} transition
  at 338.409~GHz (label 3). A marginal detection of the E line at
  338.530~GHz is indicated by label 6.  The unidentified emission
  features labeled by (?) are discussed in Sect.~\ref{sec:ch3oh}.
  }
  \label{fig:ch3oh}
\end{figure*}

\subsection{\ce{CH3OH}}
\label{sec:ch3oh}

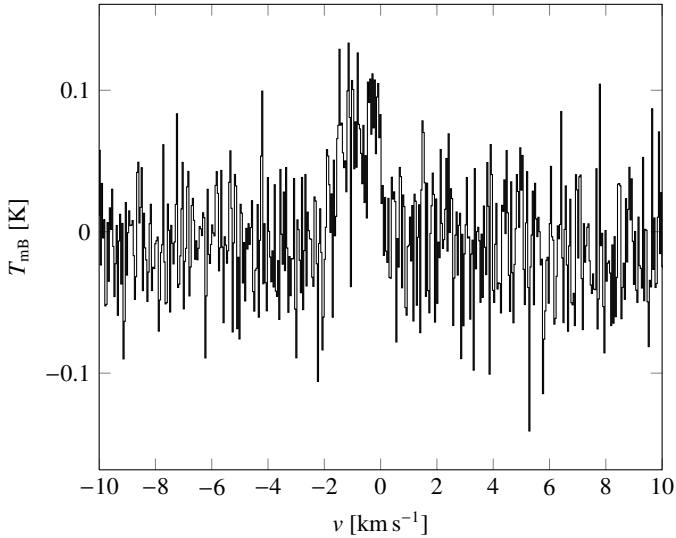
\begin{figure}
  \centering
  \begin{tikzpicture}
    \begin{axis}[xlabel={$v$ [\kms]},ylabel={$T_\mathrm{mB}$ [K]},
      const plot mark right,xmin=-10,xmax=10,
      ytick={-0.1,0,0.1}, width=\hsize]
      \addplot[mark=none] table [x index=0,y index=1]
   	{/home/miguel/project/Machholz/data/CH3OH_vel_CTS.dat};
    \end{axis}
  \end{tikzpicture}
  \caption{Averaged spectrum of the \ce{CH3OH}
  ($13_1$--$13_0$A\textsuperscript{-}) line at 342.730~GHz 
  observed on 13.20 and 16.16 UT January with the CTS.  The vertical
  axis is the calibrated main beam brightness temperature and the
  horizontal axis is the Doppler velocity in the comet rest frame.}
  \label{fig:ch3oh2}
\end{figure}

Methanol has been observed in several comets with a wide range of
abundances since its initial discovery in comet C/1989 X1 (Austin) by
\citet{1990ESASP.315..143B}.  Methanol rotational lines often appear in
multiplets at millimeter and sub-millimeter wavelengths, allowing the
estimation of the temperature and excitation conditions in the coma.
Rotational states are ordered in non-degenerate A levels and degenerate
E levels corresponding to different symmetry states.  Radiative
transitions between A and E species are strictly forbidden, and
collision-induced transitions are highly improbable.  We detected
twelve A\textsuperscript{+}, A\textsuperscript{-} and E-methanol
emission lines of the $J$ = 7--6 series in \machholz{} at frequencies
listed in Table~\ref{tbl:obs}.  In addition, there is a blend of three
emission lines at 338.513~GHz and a blend of two lines at 338.722~GHz.
Figure~\ref{fig:ch3oh} shows the observed \ce{CH3OH} rotational spectrum
between 338.10 and 338.75 GHz with labels indicating the transitions
listed in Table~\ref{tbl:obs}.  Several unidentified emission features
are found at about 338.178, 338.232, 338.277, and 338.651~GHz in the
lower sideband and 341.169, 341.543, 341.588, and 341.642~GHz in the
upper sideband, which are labeled by (?) in Fig~\ref{fig:ch3oh}. The line
widths of these features are comparable with the detected \ce{CH3OH}
lines and thus are consistent with a cometary origin, although no
plausible lines were found at those frequencies nor the corresponding
image band frequencies in the JPL Molecular Spectroscopy Catalog
\citep{1998JQSRT..60..883P} and the Cologne Database for Molecular
Spectroscopy \citep{2001A&A...370L..49M,2005JMoSt.742..215M}.
Integrated intensities and velocity shifts of the identified \ce{CH3OH}
lines with statistical uncertainties are shown in Table~\ref{tbl:obs}.
The observed velocity shifts toward the blue wing suggest asymmetric
outgassing with preferential emission in the direction toward the
observer, which should be studied with a non-spherically symmetric model.
Additionally, we detected serendipitiously a high-excitation-level
methanol transition $J = 13_1$--$13_0$A\textsuperscript{-} at
342.730~GHz in the CS/CO spectra from 13 and 16 January.
Figure~\ref{fig:ch3oh2} shows the averaged CTS spectrum from both dates.

\subsubsection{Rotational temperature}
\label{sect:rotdiag}

The rotational temperature was calculated from the relative intensities
of the individual \ce{CH3OH} lines between levels with quantum numbers $J =
7$--6 and $13_1$--$13_0$A\textsuperscript{-} using the rotation diagram
technique, assuming that the population distribution of the levels sampled by
the emission lines is in local thermodynamical equilibrium (LTE), i.e.,
described by a Maxwell-Boltzmann distribution characterized by a single
temperature.  Then the column density of the upper level within the
beam, $N_\mathrm{u}$, can be expressed as
\begin{equation} N_\mathrm{u}
  =  N \frac{g_\mathrm{u}}{Z(T_\mathrm{rot})}
  \exp\left(-\frac{E_\mathrm{u}}{k_\mathrm{B}T_\mathrm{rot}}\right),
\end{equation} 
where $g_\mathrm{u}$ is the degeneracy of the upper level, $Z$ denotes
the partition function, which is a function of temperature,
$T_\mathrm{rot}$ is the rotational temperature, $E_\mathrm{u}$ is the
energy of upper state, $k_\mathrm{B}$ represents the Boltzmann constant,
and $N$ is the total column density averaged over the beam.  This method
is commonly used in studies of the interstellar medium.  

A linear fit of $\ln(N_\mathrm{u}/g_\mathrm{u})$ versus $E_\mathrm{u}$
provides the rotational temperature as the inverse of the slope and the
averaged total column density divided by the partition function from its
intercept at $E_\mathrm{u} = 0$.  We included in the analysis the
blended transitions at 338.722 and 338.723~GHz starting from the $7_2$
and $7_{-2}$ E levels, and the 338.541 and 338.543~GHz pair from the
$7_3$~A\textsuperscript{+} and $7_3$~A\textsuperscript{-} levels, by
assigning half of the observed intensity to a virtual level of
intermediate energy \citep[see][]{1994P&SS...42..655B}.  The
high-excitation-level line at 342.730~GHz was included in the
rotational diagram analysis.  For optically thin conditions, which
normally apply to cometary lines of most volatile species, the observed
intensity, $\int T_\mathrm{mB}\,dv$, is proportional to the column
density of the upper transition level \citep{1994A&A...287..647B}:
\begin{equation} N_\mathrm{u}  =  \frac{8\pi
k_\mathrm{B}\nu_\mathrm{ul}^2}{hc^3A_\mathrm{ul}} \int
T_\mathrm{mB}\,dv, \end{equation} where $A_\mathrm{ul}$ is the Einstein
coefficient for spontaneous emission and $v_\mathrm{ul}$ the transition
frequency.

\begin{figure}
  \centering
  \begin{tikzpicture}
    \begin{semilogyaxis}[
      xlabel={$E_\mathrm{u}$ [K]},
      ylabel={$N_\mathrm{u}/g_\mathrm{u}$ [cm$^{-2}$]},
      width=\hsize, xmin = 55, xmax= 240,
      ]
      \addplot+[only marks,color=blue,error bars/.cd,y explicit,y dir=both] 
	table [x index=0,y index=1,y error index=2]
	{figures/ach3oh.txt};
      \addplot+[only marks,red,mark=o,error bars/.cd,y explicit,y dir=both] 
	table [x index=0,y index=1,y error index=2]
	{figures/ech3oh.txt};
      \addplot+[only marks,red,mark=triangle,
	every mark/.append style={rotate=180}]
	table [x index=0,y index=1]
	{figures/upperlimit.txt};
      \addplot[mark=none]
	table [x index=0,y index=1]
	{figures/rotdiagram.txt};
      \addplot[mark=none,dashed]
	table [x index=0,y index=1]
	{figures/rotdiagram_76.txt};
    \end{semilogyaxis}
  \end{tikzpicture}
  \caption{Rotation diagram for \ce{CH3OH} lines in comet \machholz{}
  including 1-$\sigma$ uncertainties.  The column density of the upper
  level divided by its degeneracy in logarithmic scale is plotted
  against the energy of the upper level.  Filled (blue) circles are
  A\textsuperscript{+} and A\textsuperscript{-}-\ce{CH3OH} lines,
  empty (red) circles denote E-\ce{CH3OH} transitions and the empty
  (red) triangle indicates the 3-$\sigma$ upper limit on the
  $J = 7_{4}$--$6_{4}$E transition at 338.530~GHz. The solid line shows
  the best linear fit with a derived rotational temperature of \rott{}.
  The dashed line shows the best linear fit excluding the $J =
  13_1$--$13_0$A\textsuperscript{-} transition at 342.730~GHz (energy of
  the upper state of $\sim$ 228 K relative to the ground state) with a
  derived rotational temperature of \rottemp.}
  \label{fig:rotdiagram}
\end{figure}
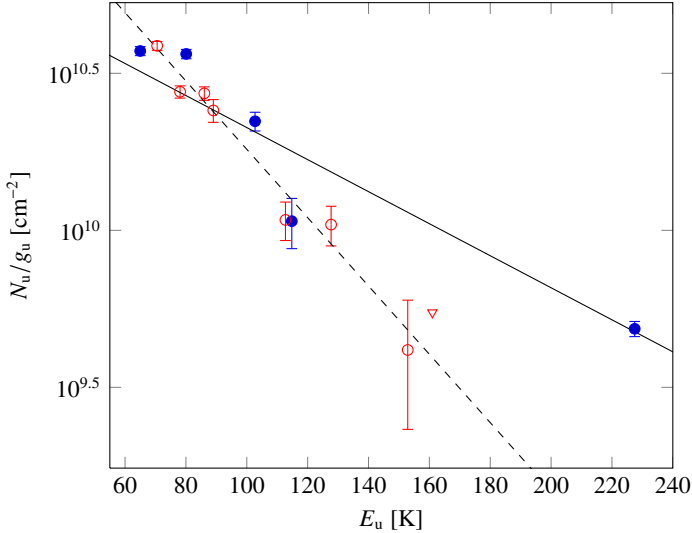

The rotation diagram for multiple \ce{CH3OH} lines that fits both A and
E-\ce{CH3OH} simultaneously was determined using a weighted linear
least-squares method where the weights are equal to the reciprocal of
the variance of each line intensity measurement.  The standard deviation
of the fitted parameters is obtained from the diagonal elements of the
covariance matrix.  Figure~\ref{fig:rotdiagram} shows the best fit for
all observed lines and only the $J$ = 7--6 series with statistical
uncertainties. In contrast to water, no reliable A/E ratio was
retrieved from radio lines for \ce{CH3OH} \citep{2007sf2a.conf..421P}.
This ratio is almost insensitive to temperature, i.e., it significantly
departs from the statistical ratio only for very low spin temperatures,
because of the small energy difference of the A and E levels.
Retrievals of the A/E ratio in \ce{CH3OH} for comet C/2001 A2 LINEAR at
infrared wavelengths are consistent with $T_\mathrm{spin} > 18\
\mathrm{K}$ \citep{2012ApJ...747...37V}.

The measured rotational temperature in \machholz{} is \rott{} with
1-$\sigma$ uncertainty from the linear best fit of the rotational
diagram that includes all the observed methanol lines on 13.20, 13.95
and 16.16 UT January (solid line in Fig.~\ref{fig:rotdiagram}). The
high-excitation line $J = 13_1$--$13_0$A\textsuperscript{-} is too
bright, indicating that the LTE assumption is not valid for that
transition.  Weighted least-squares fitting methods are known to be very
sensitive to points that are substantially farther away from the linear
relation than expected.  For a non-thermal population
distribution, a different rotational temperature is anticipated for the
$J = 13_1$--$13_0$A\textsuperscript{-} line and thus the LTE model is
not applicable to all the data.  In addition, the observations were not
simultaneous and variability in the outgassing rate is likely to be
present.  If the $J = 13_1$--$13_0$A\textsuperscript{-} line is excluded
from the rotation diagram analysis, a rotational temperature of
\rottemp{} is derived (dashed line in Fig.~\ref{fig:rotdiagram}).  

Rotational temperatures of $76 \pm 2$ K in \ce{HCN} and $93 \pm 2$ K in
\ce{H2O} were obtained on 19 January 2005 with the Near InfraRed echelle
SPECtrograph (NIRSPEC) instrument at the Keck II telescope
\citep{2009ApJ...699.1563B}.  \citet{2009ApJ...693..388K} found a
rotational temperature of $85 \pm 5$ K in \ce{H2O} and $90^{+10}_{-8}$ K
in \ce{CH4} on 30 January 2005.  Retrievals of the ortho-to-para ratios
in \ce{H2O} in \machholz{} at infrared wavelengths are consistent at
95\% confidence level with $T_\mathrm{spin} > 34\ \mathrm{K}$
\citep{2007ApJ...661L..97B} and $T_\mathrm{spin} > 27\ \mathrm{K}$
\citep{2009ApJ...693..388K}.  From the A/E/F relative abundances in
\ce{CH4} the lower limits are $T_\mathrm{spin} > 35-38\ \mathrm{K}$
\citep{2009ApJ...699.1563B} and $T_\mathrm{spin} > 27\ \mathrm{K}$
\citep{2009ApJ...693..388K}.  The nuclear spin temperature is defined as
the rotational temperature retrieved from a given spin isomers abundance
ratio assuming their respective population distributions are in LTE
conditions.  These observations at infrared wavelengths prove the
collision-dominated inner region of the coma where the population levels
are roughly in thermal equilibrium, and the rotational temperature is
expected to be similar to the kinetic temperature only in the
collisional region \citep{1994A&A...287..647B}.  However, the derived
rotational temperature of \ce{CH3OH} at 338~GHz is likely to be
different from the kinetic temperature because of relaxation of the
population levels toward fluorescence equilibrium.  We could not
determine the rotational temperature for other gas species because no
other simultaneous line observations were obtained. Therefore a kinetic
temperature of 60 K was assumed in the excitation modeling for other
species.

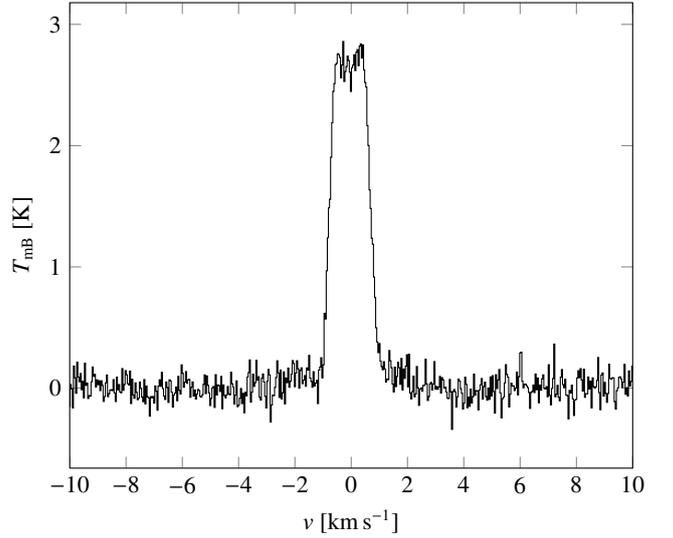
\begin{figure}
  \centering
  \begin{tikzpicture}
    \begin{axis}[xlabel={$v$ [\kms]},ylabel={$T_\mathrm{mB}$ [K]},
      const plot mark right,xmin=-10,xmax=10,
      width=\hsize]
      \addplot[mark=none] table [x index=0,y index=1]
   	{/home/miguel/project/Machholz/data/hcn_vel.dat};
    \end{axis}
  \end{tikzpicture}
  \caption{HCN (4--3) multiplet at 354.505~GHz observed with the CTS on 
  12.96~UT January with a 6420~s total on-source integration time.  The vertical
  axis is the main beam brightness temperature scaled by the beam efficiency of
  SMT and the horizontal axis is the Doppler velocity in the rest frame of
  the nucleus.
}
  \label{fig:hcn}
\end{figure}

\subsubsection{Production rate from the rotational diagram}

We derived a \ce{CH3OH} beam-averaged column density of $\langle N
\rangle / Z(T_\mathrm{rot}) = (2.22 \pm 0.45) \times 10^{11}\
\mathrm{cm}^{-2}$ from the intercept of the linear fit of the $J$ = 7--6
transitions in the rotational diagram at $E_\mathrm{u} = 0$.  Assuming
that the population of methanol levels is close to a Boltzmann
distribution, the derived column density is a good approximation of the
real distribution in the coma.  Accordingly, we omitted the transition
$J = 13_1$--$13_0$A\textsuperscript{-} in deriving the column density
for the reasons outlined in Sect.~\ref{sect:rotdiag}.  We computed the
partition function evaluated at the rotation temperature by performing a
logarithmic interpolation between the values from
\citet{2012ApJ...747...37V}.  These partition functions were computed
using two million \ce{CH3OH} levels for the range $T = 1$--500 K.
The resulting column density is $\langle N \rangle = (8.8 \pm 3.2)
\times 10^{13}\ \mathrm{cm}^{-2}$ including the 1-$\sigma$ uncertainty
in the partition function introduced by $T_\mathrm{rot}$.  A methanol
production rate of $Q_\mathrm{CH_3OH} = (6.8 \pm 2.5) \times 10^{27}~\s$
was derived from the column density assuming a Haser parent molecule
distribution with a constant expansion velocity of $\vexp$ and a
beam-pointing offset of 8\arcsec\ from the nucleus, corresponding to a
mixing ratio relative to water of $\sim 2.5$\%.  On the other hand, from
the excitation model described in Sect.~\ref{sec:radtran}, a production
rate of $Q_\mathrm{CH_3OH} = (5.5 \pm 0.6) \times 10^{27}~\s$ was
estimated from the $J$ =  7--6 transitions assuming that $T_\mathrm{kin}
= T_\mathrm{rot}$ and the same pointing offset, which corresponds to a
relative abundance relative to water of $\sim 2.1$\%. This mixing ratio
represents an intermediate value in comparison with other Oort Cloud
comets.

\subsection{HCN}

The HCN transitions are the brightest emission lines in cometary atmospheres
for ground-based submillimeter observations
\citep{1987A&A...187..475S,1994P&SS...42..655B,2010A&A...510A..55D}.
Its production rate has been found to be in a roughly constant ratio
with respect to water, therefore it is often used to derive relative molecular
abundances.  Figure~\ref{fig:hcn} shows the HCN $J$ = 4--3 rotational
line at 354.505 GHz observed by the CTS.  The line is detected with a
very high S/N of $\sim100$ after about two hours of integration.  The
intensity and velocity shift of the observed HCN emission line intensity
are shown in Table~\ref{tbl:obs}.  From the width of the line we obtain
an estimate of \vexp{} for the coma expansion velocity.  Expansion
velocities in the range $v_\mathrm{exp} = 0.5-0.8\ \kms$ are typical for
comets at this heliocentric distance and for this gas production rate
\citep{2007A&A...467..729T}.  Because HCN has a short rotational lifetime,
it is important to model its excitation including collisional and
radiative processes, which control the population levels in the outer
coma. We derive an HCN production rate of $(2.26 \pm 0.02) \times 10^{26}\
\s$ assuming the kinetic temperature to be equal to 60 K.  The water
production rate was assumed to be $Q_\ce{H2O} \sim 2.5 \times
10^{29}\ \s$, which corresponds to an HCN/\ce{H2O} mixing ratio of about
0.08\%. This value is moderately lower than the standard ratio of 0.1\%
observed at radio wavelengths for several comets over a wide range of
heliocentric distances \citep{2002EM&P...90..323B,2004come.book..391B}.

\begin{figure}
  \centering
  \begin{tikzpicture}
    \begin{axis}[xlabel={$v$ [\kms]},ylabel={$T_\mathrm{mB}$ [K]},
      const plot mark right,xmin=-10,xmax=10,
      width=\hsize]
      \addplot[mark=none] table [x index=0,y index=1]
   	{/home/miguel/project/Machholz/data/h13cn_vel.dat};
    \end{axis}
  \end{tikzpicture}
  \caption{Averaged spectrum of the \ce{H^{13}CN} (4--3) multiplet at
  345.340~GHz observed on 13.20 and 16.16 UT January with the
low-resolution AOS.  The vertical axis is the calibrated main beam
brightness temperature and the horizontal axis is the Doppler velocity
in the comet rest frame.  This line is outside the CTS and AOSC
frequency ranges.}
  \label{fig:h13cn}
\end{figure}
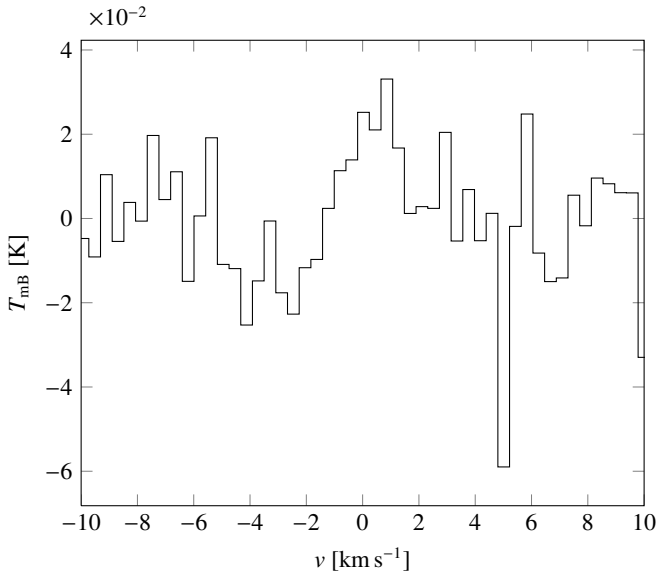

\subsection{\ce{H^{13}CN}}

Owing to the strong activity of \machholz\ and its close approach to Earth
during our observational campaign, it was possible to search for weak
transitions of rare isotopic species. The \ce{H^{13}CN} $J$ = 4--3
transition at 345.340~GHz was observed simultaneously with the CS and CO
transitions on 14.20 and 16.16 UT January with a total integration time
of $\sim 640$ minutes, and is marginally detected by averaging the
observations from both days (see Fig.~\ref{fig:h13cn}).  The inferred
production rate is $(3 \pm 1) \times 10^{24}\ \s$.  The derived
$^{12}$C/$^{13}$C isotopic ratio of \cratio{} is consistent with the
standard solar value of 89 observed in various comets.  Isotopic
fractionation is very sensitive to chemical and physical conditions
where the molecules condensed and therefore provide important clues
about how cometary material formed in the early solar system.  If comets
originate from different regions in the solar nebula, they could display
variation in their isotopic composition depending on the local
temperature.  Additional measurements of the $^{12}$C/$^{13}$C ratio in HCN
have been obtained in comets C/1995 O1 (Hale-Bopp)
\citep{1997Sci...278...90J,1999ApJ...527L..67Z} and 17P/Holmes
\citep{2008ApJ...679L..49B}, which are consistent with a solar value.
The carbon isotopic composition of several comets of different dynamical
families also agrees with a solar value from measurements of \ce{C_2}
and CH \citep[see][for reviews on the isotopic composition of
cometary volatiles]{2009EM&P..105..167J,2011IAUS..280..261B}.

\subsection{HNC}

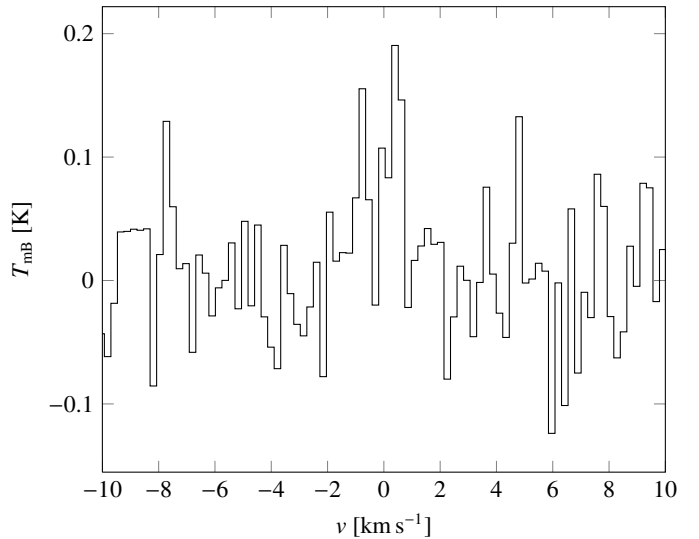
\begin{figure}
  \centering
  \begin{tikzpicture}
    \begin{axis}[xlabel={$v$ [\kms]},ylabel={$T_\mathrm{mB}$ [K]},
      const plot mark right,xmin=-10,xmax=10,
      width=\hsize]
      \addplot[mark=none] table [x index=0,y index=1]
   	{/home/miguel/project/Machholz/data/hnc_smooth_6.dat};
    \end{axis}
  \end{tikzpicture}
  \caption{Averaged spectrum of the HNC (4--3) multiplet at 362.630~GHz
  observed on 14.17 and 15.14 UT January with the CTS.  The vertical
  axis is the main beam brightness temperature and the horizontal axis
  is the Doppler velocity in the comet rest frame. The effective
  resolution after smoothing is 200 \ms. }
  \label{fig:hnc}
\end{figure}

Hydrogen isocyanide (HNC) is a metastable isomer of HCN and has been
detected in several comets with a measured  HNC/HCN ratio ranging
between 0.03--0.3 \citep{2008ApJ...675..931L}.  Figure~\ref{fig:hnc}
shows the HNC spectrum in \machholz{} observed by the CTS. A production
rate of $(7 \pm 2) \times 10^{24}\ \s$ is derived, corresponding to
a relative abundance of 3.1\% relative to HCN, a typical abundance for
moderately active comets.  The origin of cometary HNC is still a debated
topic. HNC is an abundant species that may be produced by chemical
reactions from parent volatile species  in the coma.  Models have shown
that HNC could also be formed in the coma via ion-neutral and
isomerization chemical reactions \citep{1998Natur.393..547I}.
Consequently, determining the production rate is important to
distinguish between single parent volatile and multiple precursors for
HNC.  Since we do not have precise estimates on its distribution, we
assumed the Haser formula for a parent molecule in our computation of the
production rate.

\subsection{\ce{H2CO}}

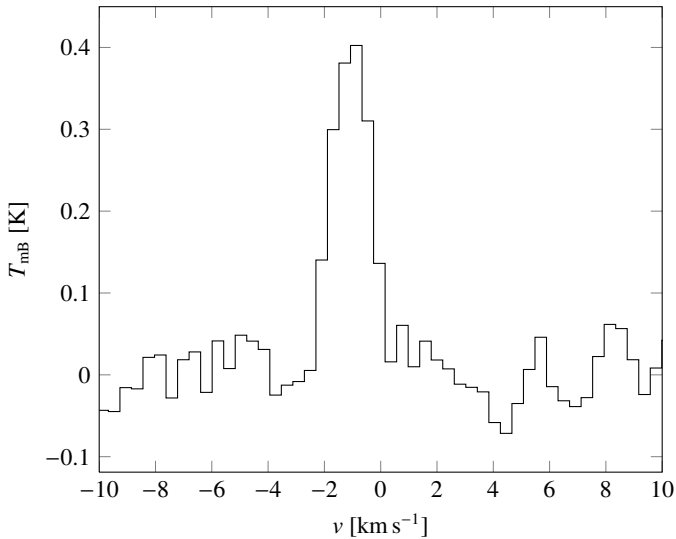
\begin{figure}
  \centering
  \begin{tikzpicture}
    \begin{axis}[xlabel={$v$ [\kms]},ylabel={$T_\mathrm{mB}$ [K]},
      const plot mark right,xmin=-10,xmax=10,
      width=\hsize]
      \addplot[mark=none] table [x index=0,y index=1]
   	{/home/miguel/project/Machholz/data/H2CO_vel_AOS.dat};
    \end{axis}
  \end{tikzpicture}
  \caption{\ce{H2CO} ($5_{15}$--$4_{14}$) transition at 351.769~GHz
  observed on 12.96 UT January with the lower resolution AOS.  The vertical
  axis is the main beam brightness temperature and the horizontal axis is the
  Doppler velocity in the comet rest frame.  This line is outside the CTS
  and AOSC frequency ranges.  The frequency scale in the range close to
  the border of the AOS can be slightly inaccurate.
  }
  \label{fig:h2co}
\end{figure}

Formaldehyde is a common molecule in the interstellar medium and has
been detected in more than 20 comets at radio wavelengths since the
first unequivocal detection at millimeter wavelengths of the
$3_{12}$–-$2_{11}$ line at 226~GHz in comet C/1989 X1 (Austin)
\citep{1992A&A...264..270C}.  Solid-phase hydrogenation reactions on the
surface of the nucleus could account for the presence of formaldehyde
and have been confirmed recently by laboratory experiments
\citep{2005ApJ...624L..29N,2009ApJ...702..291H}.  Figure~\ref{fig:h2co}
shows the \ce{H2CO} \mbox{$5_{15}$--$4_{14}$} transition at 351.769~GHz
in comet \machholz{} obtained by the AOS.  CTS and high-resolution AOS
data are not available for this transition. It is possible that the
frequency scale in this range of the AOS can be slightly inaccurate.
For the acousto-optical backends, the correspondence between channel
number and frequency is not quite linear, and the correction may be
important and ill-known on the border of the band. This line is detected
in the lower sideband of the AOS at about 300 MHz from the central
frequency with a total bandwidth of $\sim$ 1 GHz. The same \ce{H2CO}
line was observed at the Caltech Submillimeter Observatory (CSO) on
similar dates (12.1 and 13.1 January) where the velocity shift toward
the blue wing was found to be about twice smaller (Biver et al.\ in
preparation).

\ce{H2CO} has ortho- and para spin isomers that differ in the alignment
of the nuclear spin of the H atoms (parallel in the ortho state and
anti-parallel in the para state).  We assumed that cometary formaldehyde
has an ortho-to-para ratio of 3 according to the nuclear spin
statistical weights ratio.  \ce{H2CO} is a short-lived molecule and its
spatial distribution has been found to differ from that expected for a
parent molecule.  An extended source of \ce{H2CO} has been identified in
several comets \citep{1993A&A...277..677M,1994P&SS...42..655B}.
Assuming that \ce{H2CO} originates either from the nucleus or from an
extended source with a parent scale length of $L_\mathrm{p} = 8000$ km,
production rates vary between $(3.87 \pm 0.20)\times10^{26}~\s$ and $(1.09
\pm 0.06)\times10^{27}~\s$, respectively, as inferred by a model with an
8\arcsec\ average pointing offset \citep{1997PhDT........51B}.  The
first value represents a lower limit because it is expected that direct
release from the nucleus underestimates the real \ce{H2CO} production
rate.  The corresponding $Q_\ce{H2CO}/Q_\ce{H2O}$ mixing ratios are
0.14\% and 0.41\% for the parent source and extended \ce{H2CO} source
production.  These values are intermediate in comparison with the
observed mixing ratios in other Oort Cloud comets
\citep{2011IAUS..280..261B}.

\subsection{CO}

\begin{figure}
  \centering
  \begin{tikzpicture}
    \begin{axis}[xlabel={$v$ [\kms]},ylabel={$T_\mathrm{mB}$ [K]},
      const plot mark right,xmin=-10,xmax=10,ymin=-.1,ymax=.3,
      ytick={-0.1,0,0.1,0.2,0.3},
      width=\hsize]
      \addplot[mark=none] table [x index=0,y index=1]
   	{figures/CO.txt};
    \end{axis}
  \end{tikzpicture}
  \caption{CO (3--2) transition at 345.796 GHz observed on 14.94
  UT January with the CTS. The asymmetric line profile suggests a strong
  anisotropic outgassing with preferential production in the direction
  opposite to the Sun.
  }
  \label{fig:co}
\end{figure}
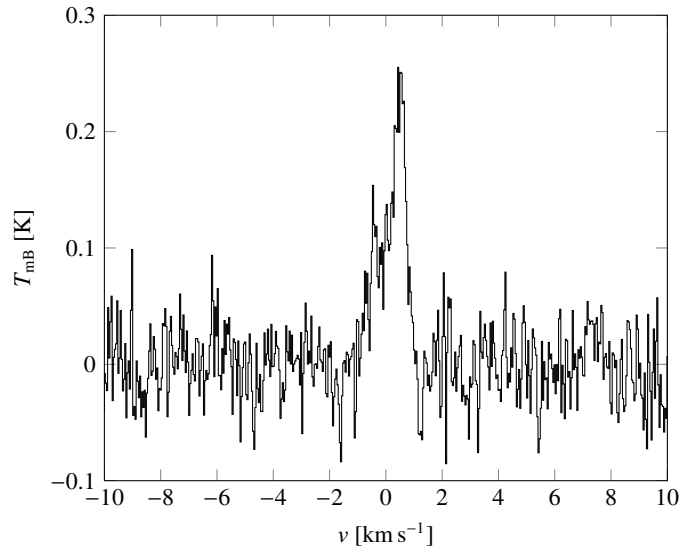

We show the carbon monoxide $J$ = 3--2 line in Fig.~\ref{fig:co}.  The
line profile is strongly asymmetric with a peak in the redshifted wing.
There is some observational evidence that the CO lines may be affected
by self-absorption effects in CO-rich and very active comets, which have
been included in our analysis \citep{2010Icar..210..898B}.  For the
observed CO transition in comet \machholz{} our model predicts an
optical depth of $\tau \sim 0.2$.  Therefore, the observed excess emission at
redshifted frequencies is attributable to kinematic effects by the
presence of a jet roughly in the antisolar direction since the solar
phase angle was small during the observations.  An anisotropic
outgassing radiative transfer model with a rotating jet-like structure
would provide a better fit to the line shape \citep[see
e.g.][]{2009A&A...505..825B}.  Based on an isotropic coma distribution
and using a kinetic gas temperature of 60 K, we deduce a CO production
rate of $(7.0 \pm 0.6) \times 10^{27}~\s$.  CO is thought to be a
parent molecule with a small additional contribution from dissociation
of \ce{CO2} and we used the Haser parent molecule distribution to
derive the CO production for simplicity.

Oort Cloud comets are expected to be enriched in hypervolatile species
such as CO and the abundance ratio of \ce{H2CO} and CO has been observed
to be correlated in several comets.  We find a relative abundance of CO
with respect to HCN of about 31 and a mixing ratio of $\sim 2.6$\%
relative to \ce{H2O} in \machholz{} .  Observations of the hypervolatile
CO in comets at infrared and radio frequencies show a wide range of
abundances \citep{2004come.book..391B}.  CO abundances observed at radio
wavelengths range between $\sim$ 0.4--20\% relative to \ce{H2O}
\citep{2011IAUS..280..261B}.  Generally, these lines are relatively weak
because of the small CO dipole, although they are the most readily
observed cometary lines at large heliocentric distances.  The CO
abundance in eight Oort Cloud comets observed by infrared ground-based
spectroscopy ranged between $\sim$ 1--20\% relative to water
\citep{2003AdSpR..31.2563M}.  Thus the measured $Q_\ce{CO}/Q_\ce{H2O}$
mixing ratio lies toward the low end of the observed range in Oort
Cloud comets.  Depletion of CO would therefore suggest that \machholz{}
has been injected into the inner solar system before and is not a
dynamically new comet.

\subsection{CS}

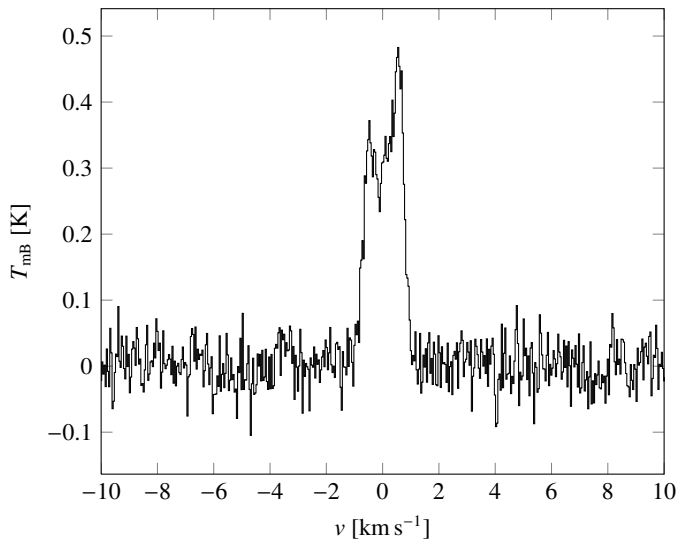
\begin{figure}
  \centering
  \begin{tikzpicture}
    \begin{axis}[xlabel={$v$ [\kms]},ylabel={$T_\mathrm{mB}$ [K]},
      const plot mark right,xmin=-10,xmax=10,
      width=\hsize]
      \addplot[mark=none] table [x index=0,y index=1]
   	{/home/miguel/project/Machholz/data/cs_vel.dat};
    \end{axis}
  \end{tikzpicture}
  \caption{CS (7--6) emission line at 342.883 GHz observed on
  14.94 UT January with the CTS.  A double-peaked profile is observed.  The
  velocity scale in the horizontal axis is in the comet rest frame.
  }
\label{fig:cs}
\end{figure}

The CS radical is believed to be produced in cometary coma from carbon
disulfide (\ce{CS2}) and has been observed in several comets in the UV
and radio wavelengths \citep{2004come.book..425F,2004come.book..391B}.
Figure~\ref{fig:cs} shows the observed CS $J$ = 7--6 emission line at
342.883~GHz.  The line profile is asymmetric with a peak in the
redshifted side that resembles the \ce{CO} $J$ = 3--2 line shape shown in
Fig.~\ref{fig:co}.  This suggests an anisotropic outgassing in the
anti-sunward direction. It is expected that the line is optically thin
and absorption effects in the foreground are negligible according to our
model.  To obtain the density profile we used a photodissociation rate of
$\beta_\mathrm{CS} = 2.5 \times 10^{-5} \s$ at $\rh = 1\ \mathrm{AU}$
derived from spectroscopic observations
\citep{2007A&A...475.1131B,2011A&A...528A.142B}.  This photodissociation
rate corresponds to a scale-length of $28\,000$ km, assuming an expansion
velocity of $\vexp{}$ obtained from the width of the HCN line.  A CS
production rate of $(1.15 \pm 0.04) \times 10^{26}\ \s$ is estimated from
a spherically symmetric model with direct release from the nucleus.
Assuming a distributed source in the coma with scale length
$L_\mathrm{p} = 650~\text{km}$, which corresponds to production from
\ce{CS2} \citep{2007A&A...475.1131B}, a production rate of $(1.29 \pm
0.04) \times10^{26}\ \s$ is inferred. The derived mixing ratio with
respect to water of $\sim$ 0.04\% is lower than the typical value
measured in other comets at radio wavelengths
\citep{2011IAUS..280..261B,2009EM&P..105..267C}.

\subsection{Short-term outgassing variability}

Several comets have shown periodic variations in the outgassing that
are associated with the rotation of the nucleus, combined with a
long-term seasonal variation with a peak emission close to perihelion.
In some cases, the rotation periods of cometary nuclei have been
determined by measuring this periodic outgassing variability using
various techniques \citep[see][for a review]{2004come.book..281S}.
Rotation periods have been derived from the periodic variability of the
HCN production rate in comets 9P/Tempel 1 \citep{2007Icar..187..253B},
73P-C/Schwassmann-Wachmann 3 \citep{2010A&A...510A..55D} and 2P/Encke
\citep{2011Icar..215..153J}.  To address the question of short-term
periodic variability in the production rates of \machholz{}, we selected
the HCN and CS observations from one night, 12--13 January, which have
sufficient S/N.

\begin{figure}
  \centering
  \begin{tikzpicture}
    \begin{axis}[xlabel={Date [UT]},ylabel={$Q_\mathrm{HCN}$ [$10^{26}\ \s$]},
      width=\hsize]
      \addplot+[only marks,mark=o,color=black]
	plot[error bars/.cd,y dir=both,y explicit]
	table [x index=0,y index=1, y error index=2]
   	{/home/miguel/project/Machholz/data/hcnvariability.dat};
    \end{axis}
  \end{tikzpicture}
  \caption{HCN production rates in comet \machholz{} as a function of time on
  12-13 January 2005.  Each data point represents an 8-minute
  observation.
  }
  \label{fig:hcnvariability}
\end{figure}
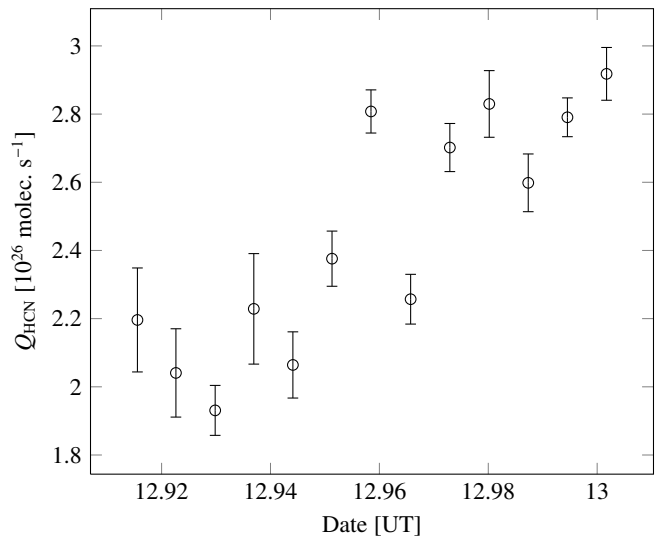

The production rates for each individual scan of the HCN observations
are shown in Fig.~\ref{fig:hcnvariability}. The HCN $J$ = 4--3 line was
detected in each single 8-minute scan with an S/N of $\sim 15$.  We
observe an increase of the production rate by roughly 40\% during the
2-hour observing interval with a mean value of $(2.09 \pm 0.02) \times
10^{26}\ \s$, although the first scans at the beginning of the observations were
obtained at lower elevation starting at 45\degr\ and could be affected
by stability problems in the instrument or changes in the main beam
efficiency.  The HCN evolution suggests a periodic variation in the
activity of the comet potentially induced by the rotation of the nucleus
due to the presence of active regions on the surface.  The large
amplitude of the variation can be explained by an intrinsic activity
change as one or more active areas are exposed to solar radiation during
the rotation cycle.  This variation is compatible with the rotation
periods of $9.1 \pm 0.2$ hours derived by \citet{2009A&A...494..379R}
from broadband filter optical photometry, $9.1 \pm 1.9$ hours reported
by \citet{2005IAUC.8480....3S} from studying dust fans visible in
$R$-band images, $17.60 \pm 0.05$ hours from narrowband CN images
reported by \citet{2007AJ....133.2001F}, and $17.8 \pm 0.5$ hours by
determining the inner coma morphology from photometric variations
\citep{2012Ap&SS.337..531M}.  However, we are not able to perform a
detailed periodicity analysis of the HCN production curve due to the
insufficient temporal coverage because the total time span of our
observations is less than one night.  Hence it is not possible to
determine if the brightness increase is a periodic phenomenon or due to
non-periodic activity changes.

\begin{figure}
  \centering
  \begin{tikzpicture}
    \begin{axis}[xlabel={Date [UT]},ylabel={$Q_\mathrm{CS}$ [$10^{26}\ \s$]},
      width=\hsize]
      \addplot+[only marks,mark=o,color=black]
	plot[error bars/.cd,y dir=both,y explicit]
	table [x index=0,y index=1, y error index=2]
   	{/home/miguel/project/Machholz/data/CLASS_CO(3-2)_CTSA.dat};
    \end{axis}
  \end{tikzpicture}
  \caption{Time evolution of the retrieved CS production rates in comet
  \machholz{} on 13 January 2005.  Each data point represents an 8-minute
  observation with statistical uncertainties.
  }
  \label{fig:csvariability}
\end{figure}
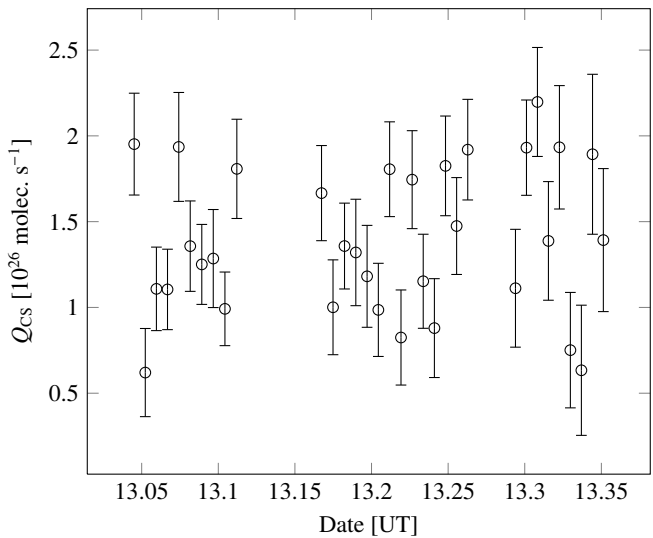

Figure~\ref{fig:csvariability} shows the evolution of the CS outgassing rate on
13 January during a period of 7.5 hours. The CS line is relatively strong and
has one of the highest peak antenna temperatures after the HCN line.
CS is detected in most of the individual 8-minute scans with an S/N between 3-6.
The comet displayed a strong variability with a 40\% variation around the mean
value.  It is likely that this brightness variability is not a periodic
event.  Small outbursts or activity changes are expected as comets
approach the Sun, which could explain the observed variability
\citep{2005Sci...310..258A}.  It is also possible that the strong
fluctuations in the CS production could be partly explained by pointing
uncertainties that affect the derived production rates combined with
instrument instabilities.
Observations of the CS line on 16 January have a lower S/N and only in a
fraction of the individual scans the line is detected with at least a
3-$\sigma$ confidence level.

\section{Discussion}\label{sec:discussion}

We have observed several molecular species in comet \machholz{} when it
was close to perihelion in the sub-millimeter wavelength range with the
SMT during six consecutive nights in January 2005.  These observations
led to the detection of several \detected{} rotational lines.  Our main
goal was to measure the relative production rates of several parent and
daughter volatiles in this comet and estimate the \ce{CH3OH} rotational
temperature.  We derive an HCN production rate of $(2.26 \pm 0.03)
\times10^{26}\ \s$ at heliocentric distance of 1.2 AU, which corresponds
to a $Q_\ce{HCN}/Q_\ce{H2O} \sim 0.08$\%, using a spherically symmetric
radiative transfer numerical code that includes collisional effects
between neutrals and electrons  and radiative pumping of the fundamental
vibrational levels by solar radiation \citep{1989A&A...216..278B}.  In
addition to the statistical noise, the precision of the production rates
is affected by the limited pointing accuracy at the time the
observations were performed. An 8\arcsec\ pointing offset was
included in the computation of the production rates, which results in
$\sim$  20\% uncertainty in the derived values.  Mixing ratios relative
to hydrogen cyanide and water are listed in Table~\ref{tbl:mixing} for
the production rates derived from the weighted average value of the
observations on different dates during our campaign.  The asymmetric
shape of the CO and CS line profiles suggests that there is preferential
outgassing from the anti-sunward side of the nucleus, while the HCN and
\ce{H2CO} line profiles are fairly symmetric, although the latter is
shifted toward the blue wing perhaps due to an instrumental effect.
Using the rotational diagram technique, we retrieve a cold rotational
temperature of \rottemp\ for the methanol energy levels sampled by the
$J$ = 7--6 transitions.

Retrieved molecular abundances relative to water are comparable to those
obtained in Oort Cloud comets
\citep{2002EM&P...90..323B,2009P&SS...57.1162C,2011ApJ...734L...8D}.
Our $Q_\ce{HCN}/Q_\ce{H2O}$ mixing ratio of $\sim 0.084$\% is slightly
lower than the typical value of 0.1\% observed at radio wavelengths, and
50\% lower than those found by \citet{2009ApJ...699.1563B} from averaged
28 November 2004 and 19 January 2005 observations and by
\citep{2009ApJ...703..121K} from observations on 30 January 2005 using
the same instrument.  The observed difference between our measurement of
the HCN mixing ratio and those derived at infrared wavelengths is fairly
typical.  The mixing ratio of \ce{CH3OH} relative to \ce{H2O} agrees
with the infrared measurement in comet \machholz{} by
\citet{2009ApJ...699.1563B} performed on 19 January 2005 at $\rh =
1.208$ AU that is closer in time with our observations. Considering the
pointing uncertainty that introduces an error of $\sim$ 20\%, our
measurement is consistent within 1-$\sigma$ with the revised abundances
retrieved from the infrared observations on 19 and 30 January 2005 by
\citet{2012ApJ...747...37V} using a new line-by-line model for the
$\nu_3$ fundamental band of \ce{CH3OH}.  The averaged CO production rate
measured on 13 and 16 January agrees with that obtained by
\citet{2009ApJ...699.1563B} on 29 November 2004 at $\rh = 1.493$ AU
($(6.3 \pm 0.3)\times10^{27} \s$) within confidence limits.  In
contrast, the $Q_\ce{CO}/Q_\ce{H2O}$ mixing ratio of $\sim 2.6$\% is
almost a factor of two lower than that derived by
\citet{2009ApJ...699.1563B}.  For a parent molecule distribution, the
derived $Q_\ce{H2CO}/Q_\ce{H2O}$ ratio is 0.14\%, and
$Q_\ce{H2CO}/Q_\ce{HCN}$ is 1.7\%, which is more reliable since these
molecules were observed simultaneously.  These values are intermediate
between those measured by \citet{2009ApJ...699.1563B} and
\citet{2009ApJ...703..121K}. On the other hand, the values inferred from
a daughter molecule extended source distribution with a scale length of
$L_\mathrm{p} = 8000$ km are about a factor of 3 higher.  This
illustrates that the derived production rates depend strongly
on the assumed scale length of the parent molecule. This is particularly
the case for \ce{H2CO} where the beam size is slightly smaller than
$L_\mathrm{p}$, so it may be possible that this value overestimates the
production rate.  An HNC abundance of 3.1\% with respect to HCN is found
assuming direct release from the nucleus with a Haser distribution
(3.0\% if infrared pumping of the fundamental vibrational levels is not
considered), which suggests that this molecule may be destroyed by
chemical reactions.  This value is compatible with those observed in
other comets given the dependence on heliocentric distance -- except
73P/Schwassmann-Wachmann and the very active Hale-Bopp (C/1995 O1)
\citep[see][]{2008ApJ...675..931L}.

Outgassing variations induced by the nucleus rotation are expected to
appear from non-sphericity of the nucleus or the presence of active
region areas on the surface.  The variability observed in the HCN and CS
outgassing rates on 12--13 January is affected by inaccurate pointing,
which introduces an uncertainty of about the expected variation caused
by the rotation of the nucleus.  HCN production rates show a uniform
brightness increase over a period of two hours, which could be a periodic
phenomenon consistent with the rotation period of the nucleus derived
using different methods
\citep{2005IAUC.8480....3S,2007AJ....133.2001F,2009A&A...494..379R,2012Ap&SS.337..531M},
but our observations do not provide sufficient phase coverage to
constrain the rotation period.

Comets are the most pristine objects in the solar system and have not
undergone substantial thermal processing.  Cometary ices are more
sensitive to thermal processing than dust. Hence their abundances
provide indications about the formation and evolution of material in the
early solar nebula. Dynamically new comets are expected to be enriched
in volatile species, and CO is the most volatile component observed in
\machholz{}. However, our observations show a relative depletion of CO
and and an intermediate-range mixing ratio of \ce{H2CO} and \ce{CH3OH}
compared to the range of values measured in other comets
\citep{2002EM&P...90..323B}.  It is believed that the composition of icy
material on the surface of the nucleus is altered by the exposure to
solar radiation.  Thus, these observations provide a hint about its
thermal/dynamical history and suggest that the comet has visited the
inner solar system previously.  Formation regions of comets of Oort
Cloud comets vary from 5--30 AU from the protosun according to the
standard ``Nice model'' scenario \citep{2008ssbn.book..275M} and can be
constrained by observations of the chemical composition.  The relative
depletion of several volatiles in \machholz{} agrees with the
determination of the formation region in the inner region of the solar
nebula compared with other Oort Cloud comets, derived from the lower
limit of the retrieved nuclear spin temperatures and dynamical models of
the evolution of planetesimals in the solar system
\citep{2009ApJ...703..121K,2009ApJ...693..388K}.

Submillimeter spectroscopy of cometary atmospheres is a useful approach
for studying the diversity of species that sublimate when a comet approaches
the Sun.  These observations demonstrate the capabilities of the new CTS
based on digital technology installed at the SMT to support solar system
observation programs, further established with the cometary observations
presented in
\citet{2004DPS....36.2505K,2004ExA....18...77V,2010A&A...510A..55D,2010ApJ...715.1258P,2011Icar..215..153J},
and observations of the Venusian mesosphere in
\citet{2008P&SS...56.1688R,2008P&SS...56.1368R}.  Our observations of
comet \machholz{} were analyzed for noise properties, standing wave and
spectral line properties as compared to model predictions.  The
spectrometer is found to perform according to its original specification
in terms of sensitivity, spectral resolution and stability.

\begin{acknowledgements}
The SMT is operated by the Arizona Radio Observatory (ARO), Steward
Observatory, University of Arizona.  We are grateful to the ARO staff
for their support during these observations.  This work was supported by
the Special Priority Program 1488 of the German Science Foundation.
M.dV.B.\ acknowledges fruitful discussions with Michal~Drahus during the
course of this work.  We thank the referee, Michael~F.~A'Hearn, for
helpful comments that improved the manuscript.
\end{acknowledgements}

\bibliographystyle{aa}
\bibliography{ads,preprints}

\begin{landscape}
\begin{table}
  \caption{Rotational emission lines in comet \machholz{} observed by SMT on
  13--16 January 2005.}
  \label{tbl:obs}
  \begin{center}
  \begin{tabular}{c c c r@{}l c c r@{}l c c r}
    \hline\hline
    Molecule & Label\tablefootmark{a} & Date\tablefootmark{b} &
    \multicolumn{2}{c}{Transition} &Frequency\tablefootmark{c} &
    $\int T_\textrm{mB}\, dv$ & 
    \multicolumn{2}{c}{Velocity shift\tablefootmark{d}} & $Q$\tablefootmark{e} & $Q$\tablefootmark{f} &
    \multicolumn{1}{c}{Uncertainty\tablefootmark{g}} \\
    & & (UT) & & & (GHz) & (K \kms) & \multicolumn{2}{c}{(\ms)} &
    ($\s$) & ($\s$) & \multicolumn{1}{c}{($\s$)}\\
    \hline
HCN\tablefootmark{h}          &                             & 12.96                   & 4--        & 3                            & 354.505476                    & $3.962 \pm 0.033$                  & $-31\,\pm\,$                   &  7                           &  $(1.87 \pm 0.02) \times 10^{26}$                   & $(2.26 \pm 0.02) \times 10^{26}$                   & $0.34 \times 10^{26}$                   \\
H$_2$CO\tablefootmark{i}      &                             & 12.96                   & $5_{15}$-- & $4_{14}$                     & 351.768645                    & $0.675 \pm 0.035$                  & $-783\,\pm\,$                  &  65                          &  $(9.64 \pm 0.50) \times 10^{26}$                   & $(1.09 \pm 0.06) \times 10^{27}$                   & $1.64 \times 10^{26}$                   \\
CO                            &                             & 13.20                   & 3--        & 2                            & 345.795989                    & $0.194 \pm 0.012$                  & $200\,\pm\,$                   &  38                          &  $(5.66 \pm 0.35) \times 10^{27}$                   & $(6.81 \pm 0.42) \times 10^{27}$                   & $1.02 \times 10^{27}$                    \\
CS\tablefootmark{i}           &                             & 13.20                   & 7--        & 6                            & 342.882850                    & $0.550 \pm 0.010$                  & $-50\,\pm\,$                   &  29                          &  $(1.15 \pm 0.02) \times 10^{26}$                   & $(1.37 \pm 0.03) \times 10^{26}$                   & $0.17 \times 10^{26}$                   \\
\ce{CH3OH}                    &                             & 13.20                   & $13_1$--   & $13_{0}$A\textsuperscript{-} & 342.729796                    & $0.212 \pm 0.010$                  & $-826\,\pm\,$                  &  28                          &  $(5.84 \pm 0.28) \times 10^{27}$                   & $(7.07 \pm 0.33) \times 10^{27}$                   & $1.1 \times 10^{27}$                   \\
\ce{CH3OH}                    & \ (1)\ \rdelim\}{17}{0mm}[] & \multirow{17}{*}{13.95} & $7_{0}$--  & $6_{0}$E                     & 338.124488                    & $0.468 \pm 0.021$                  & $-358\,\pm\,$                  &  97 \ \ \rdelim\}{17}{0mm}[] &  \multirow{17}{*}{$(4.55 \pm 0.47) \times 10^{27}$} & \multirow{17}{*}{$(5.48 \pm 0.56) \times 10^{27}$} & \multirow{17}{*}{$0.82 \times 10^{27}$} \\
\ce{CH3OH}                    & (2)                         &                         & $7_{-1}$-- & $6_{-1}$E                    & 338.344588                    & $0.644 \pm 0.021$                  & $-101\,\pm\,$                  &  21                          \\
\ce{CH3OH}                    & (3)                         &                         & $7_{0}$--  & $6_{0}$A\textsuperscript{+}  & 338.408698                    & $0.633 \pm 0.021$                  & $-41\,\pm\,$                   &  93                          \\
\ce{CH3OH}                    & (4)                         &                         & $7_{-4}$-- & $6_{-4}$E                    & 338.504065                    & $0.048 \pm 0.021$                  & $-58\,\pm\,$                   &  93                          \\
\ce{CH3OH}                    &                             &                         & $7_{4}$--  & $6_{4}$A\textsuperscript{+}  & 338.512632\rdelim\}{3}{0mm}[] &                                    &                                \\
\ce{CH3OH}                    & (5)                         &                         & $7_{4}$--  & $6_{4}$A\textsuperscript{-}  & 338.512644                    & $0.313 \pm 0.021$                  & $-114\,\pm\,$                  &  55                          \\
\ce{CH3OH}                    &                             &                         & $7_{2}$--  & $6_{2}$A\textsuperscript{-}  & 338.512853                    &                                    &                                \\
\ce{CH3OH}                    & (6)                         &                         & $7_{4}$--  & $6_{4}$E                     & 338.530257                    & $0.037 \pm 0.021$                  & $-162\,\pm\,$                  &  132                         \\
\ce{CH3OH}                    & (7)                         &                         & $7_{3}$--  & $6_{3}$A\textsuperscript{+}  & 338.540826\rdelim\}{2}{0mm}[] & \multirow{2}{*}{$0.296 \pm 0.027$} & \multirow{2}{*}{$-48\,\pm\,$}  &  \multirow{2}{*}{98}         \\
\ce{CH3OH}                    & (8)                         &                         & $7_{3}$--  & $6_{3}$A\textsuperscript{-}  & 338.543152                    &                                    &                                \\
\ce{CH3OH}                    & (9)                         &                         & $7_{-3}$-- & $6_{-3}$E                    & 338.559963                    & $0.146 \pm 0.021$                  & $-61\,\pm\,$                   &  335                         \\
\ce{CH3OH}                    & (10)                        &                         & $7_{3}$--  & $6_{3}$E                     & 338.583216                    & $0.150 \pm 0.021$                  & $-300\,\pm\,$                  &  144                         \\
\ce{CH3OH}                    & (11)                        &                         & $7_{1}$--  & $6_{1}$E                     & 338.614936                    & $0.466 \pm 0.023$                  & $-353\,\pm\,$                  &  40                          \\
\ce{CH3OH}                    & (12)                        &                         & $7_{2}$--  & $6_2$A\textsuperscript{+}    & 338.639802                    & $0.349 \pm 0.024$                  & $-212\,\pm\,$                  &  70                          \\
\ce{CH3OH}                    & \multirow{2}{*}{(13)}       &                         & $7_{2}$--  & $6_{2}$E                     & 338.721693\rdelim\}{2}{0mm}[] & \multirow{2}{*}{$0.744 \pm 0.031$} & \multirow{2}{*}{$-389\,\pm\,$} &  \multirow{2}{*}{55}         \\
\ce{CH3OH}                    &                             &                         & $7_{-2}$-- & $6_{-2}$E                    & 338.722898                    &                                    &                                \\
\ce{CH3OH}                    & (14)                        &                         & $7_{1}$--  & $6_{1}$A\textsuperscript{-}  & 341.415615                    & $0.613 \pm 0.021$                  & $-374\,\pm\,$                  &  29                          \\
HNC\tablefootmark{h}          &                             & 14.17                   & 4--        & 3                            & 362.630303                    & $0.100 \pm 0.039$                  & $457\,\pm\,$                   &  303                         &  $(4.2 \pm 1.6) \times 10^{24}$                     & $(5.2 \pm 1.8) \times 10^{24}$                     & $0.7 \times 10^{24}$                  \\
H$^{13}$CN\tablefootmark{h,j} &                             & 14.94                   & 4--        & 3                            & 345.339759                    & $0.050 \pm 0.018$                  &                                &                              &  $(2.5 \pm 0.9) \times 10^{24}$                   & $(3 \pm 1) \times 10^{24}$                   & $0.5 \times 10^{24}$                  \\
HNC\tablefootmark{h}          &                             & 15.14                   & 4--        & 3                            & 362.630303                    & $0.170 \pm 0.039$                  & $20\,\pm\,$                    &  137                         &  $(7.1 \pm 1.6) \times 10^{24}$                     & $(8.8 \pm 2.0) \times 10^{24}$                     & $1.3 \times 10^{24}$                  \\
CO                            &                             & 16.16                   & 3--        & 2                            & 345.795989                    & $0.205 \pm 0.022$                  & $119\,\pm\,$                   &  64                          &  $(5.98 \pm 0.64) \times 10^{27}$                   & $(7.20 \pm 0.77) \times 10^{27}$                   & $1.08 \times 10^{27}$                    \\
CS\tablefootmark{i}           &                             & 16.16                   & 7--        & 6                            & 342.882850                    & $0.492 \pm 0.024$                  & $-50\,\pm\,$                   &  29                          &  $(1.03 \pm 0.05) \times 10^{26}$                   & $(1.22 \pm 0.06) \times 10^{26}$                   & $0.18 \times 10^{26}$                   \\
\ce{CH3OH}                    &                             & 16.16                   & $13_1$--   & $13_{0}$A\textsuperscript{-} & 342.729796                    & $0.155 \pm 0.026$                  & $-90\,\pm\,$                   &  93                          &  $(4.27 \pm 0.72) \times 10^{27}$                   & $(5.17 \pm 0.87) \times 10^{27}$                   & $0.78 \times 10^{27}$                   \\

    \hline
  \end{tabular}
  \end{center}
  \tablefoot{Observed flux densities integrated over velocity, Doppler
  velocity shifts and production rates derived using the comet model
  described in Sect.~\ref{sec:radtran} are shown with statistical
  uncertainties.}
  \tablefoottext{a}{Labels in Fig.~\ref{fig:ch3oh}}
  \tablefoottext{b}{Mid-time of the observations recorded as fractional
  days in UT.}
  \tablefoottext{c}{The line frequencies were obtained from the latest
  online edition of the JPL Molecular Spectroscopy Catalog
  \citep{1998JQSRT..60..883P}.}
  \tablefoottext{d}{The velocity offsets are computed with respect to
  the optocenters of the complete components for blended lines.}
  \tablefoottext{e}{Production rates assuming a  
  pointing offset of 2\arcsec.}
  \tablefoottext{f}{Production rates assuming a
  pointing offset of 8\arcsec{}.}
  \tablefoottext{g}{The uncertainty in the production rates are 1-$\sigma$
  uncertainties considering the receiver sideband gain ratio, main beam
  efficiency and kinetic temperature errors.}
  \tablefoottext{h}{Line intensities are the sum of the hyperfine
  components.}
  \tablefoottext{i}{Production rates were derived assuming a
  daughter product extended source distribution with a scale length of
  $L_\mathrm{p} = 8000$ km for \ce{H2CO} and $L_\mathrm{p} = 650$ km for
  CS.}
  \tablefoottext{j}{Weighted average of observations obtained on 13.20
  and 16.16 UT January.}
\end{table}
\end{landscape}

\end{document}